\documentclass[prx,twocolumn,aps,letterpaper, preprintnumbers,superscriptaddress]{revtex4-2}
\usepackage{amsthm,amsmath,amssymb}

\usepackage{mathrsfs}
\usepackage{amsthm}
\usepackage{amsmath}
\usepackage{amssymb}
\usepackage{graphicx}
\usepackage{epstopdf}
\usepackage{url}
\usepackage{subfigure}
\usepackage{float}
\usepackage{bm}
\usepackage{ifthen}
\usepackage[usenames,dvipsnames]{color}
\usepackage{mathrsfs}
\usepackage[colorlinks=true,citecolor=blue,urlcolor=black]{hyperref}

\newcommand{\tn}[1]{\textnormal{#1}}
\newcommand{\be}{\begin{equation}}
\newcommand{\ee}{\end{equation}}

\newcommand{\esec}{\eps_{\tn{sec}}}

\newcommand{\sket}[1]{{\ensuremath{\lvert#1\rangle}}}
\newcommand{\lket}[1]{{\ensuremath{\left\lvert#1\right\rangle}}}
\newcommand{\ket}[1]{\if@display\lket{#1}\else\sket{#1}\fi}

\newcommand{\sbra}[1]{{\ensuremath{\langle#1\rvert}}}
\newcommand{\lbra}[1]{{\ensuremath{\left\langle#1\right\rvert}}}
\newcommand{\bra}[1]{\if@display\lbra{#1}\else\sbra{#1}\fi}

\newcommand{\sbraket}[2]{{\ensuremath{\langle#1\rvert#2\rangle}}}
\newcommand{\lbraket}[2]{{\ensuremath{\left\langle#1\!\left\rvert\vphantom{#1}#2\right.\!\right\rangle}}}
\newcommand{\braket}[2]{\if@display\lbraket{#1}{#2}\else\sbraket{#1}{#2}\fi}

\newcommand{\sketbra}[2]{{\ensuremath{\lvert #1\rangle\!\langle #2\rvert}}}
\newcommand{\lketbra}[2]{{\ensuremath{\left\lvert #1\right\rangle\!\!\left\langle #2\right\rvert}}}
\newcommand{\ketbra}[2]{\if@display\lketbra{#1}{#2}\else\sketbra{#1}{#2}\fi}

\newcommand{\eps}{\varepsilon}

\newcommand{\cK}{\mathcal{K}}

\newcommand{\cZ}{\mathcal{Z}}

\newcommand{\rs}{{\rm{1}}}
\newcommand{\rd}{{\rm{2}}}
\newcommand{\rdd}{{\rm{3}}}

\theoremstyle{plain}

\newtheorem{theorem}{Theorem}

\newtheorem{proposition}[theorem]{Proposition}
\theoremstyle{definition}
\newtheorem{definition}{Definition}

\begin{document}

\date{\today}

\title{One-Time Universal Hashing Quantum Digital Signatures without Perfect Keys}
	
\author{Bing-Hong Li}	
\author{Yuan-Mei Xie}
\author{Xiao-Yu Cao}
\author{Chen-Long Li}
\affiliation{National Laboratory of Solid State Microstructures and School of Physics, Collaborative Innovation Center of Advanced Microstructures, Nanjing University, Nanjing 210093, China}
\author{Yao Fu}\email{yfu@iphy.ac.cn}
\affiliation{Beijing National Laboratory for Condensed Matter Physics and Institute of Physics, Chinese Academy of Sciences, Beijing 100190, China}
\author{Hua-Lei Yin}\email{hlyin@nju.edu.cn}
\author{Zeng-Bing Chen}\email{zbchen@nju.edu.cn}
\affiliation{National Laboratory of Solid State Microstructures and School of Physics, Collaborative Innovation Center of Advanced Microstructures, Nanjing University, Nanjing 210093, China}	 
\date{\today}

\begin{abstract}
Quantum digital signatures (QDS), generating correlated bit strings among three remote parties for signatures through quantum law, can guarantee non-repudiation, authenticity, and integrity of messages. 
Recently, one-time universal hashing QDS framework, exploiting the quantum asymmetric encryption and universal hash functions, has been proposed to significantly improve the signature rate and ensure unconditional security by directly signing the hash value of long messages. However, similar to quantum key distribution, this framework utilizes keys with perfect secrecy by performing privacy amplification that introduces cumbersome matrix operations, thereby consuming large computational resources, causing delays, and increasing failure probability. 
Here, we prove that, different from private communication, imperfect quantum keys with partial information leakage can be used for digital signatures and authentication without compromising the security while having eight orders of magnitude improvement on signature rate for signing a megabit message compared with conventional single-bit schemes. This study significantly reduces the  delay for data postprocessing and is compatible with any quantum key generation protocols. In our simulation, taking two-photon twin-field key generation protocol as an example, QDS can be practically implemented over a fiber distance of 650 km between the signer and receiver. For the first time, this study offers a cryptographic application of quantum keys with imperfect secrecy and paves a way for the practical and agile implementation of digital signatures in a future quantum network.
\end{abstract}

\maketitle

%%%%%%%%%%%%%%%%%%%%%%%%%%  body  %%%%%%%%%%%%%%%%%%%%%%%%%%
\section{Introduction}
Digital signatures are cryptographic primitives that offer data authenticity and integrity~\cite{diffie1976new}, especially for the non-repudiation of sensitive information.
It has become an indispensable and essential technique in the global internet owing to its wide application especially in digital financial transactions, email, and digital currency. However, the security of classical digital signatures, guaranteed by public-key infrastructure~\cite{demillo1978foundations,rivest1978method,elgamal1985public}, is threatened by rapidly developing algorithms~\cite{Stevens:2017:The,boudot2020comparing} and quantum computing~\cite{shor1994algorithms}.
Different from classical digital signatures, quantum digital signatures (QDSs) can provide a higher level of security, information-theoretic security, by employing the fundamental principles of quantum mechanics. That is, QDS can protect data integrity, authenticity, and non-repudiation even if the attacker utilizes unlimited computational power.
The rudiment of the single-bit QDS scheme was introduced in 2001~\cite{gottesman2001quantum}, but it could not be implemented due to some impractical requirements such as high-dimensional single-photon states and quantum memories.
Subsequently, there have been many developments to remove these impractical requirements~\cite{clarke2012experimental,Collins:2014:Realization,dunjko2014quantum}, making QDS closer to real implementation.
Furthermore, based on non-orthogonal encoding~\cite{Yin:practical:2016} and
orthogonal encoding~\cite{Amiri:Secure:2016}, respectively, two independent single-bit QDS protocols without secure quantum channels were proposed and proved to be secure for the first time.
These two protocols have triggered numerous achievements of single-bit QDS theoretically~\cite{puthoor2016measurement,PhysRevA.94.042314,yang2017theoretically,Thornton:2019:CV,Qu:19,zhang2020practical,lu2021efficient,zhang2021twin,PhysRevA.103.012410,Weng:21,qin2022quantum,zhang2022practical} and experimentally~\cite{PhysRevA.95.032334,collins2017experimental,PhysRevA.95.042338,roberts2017experimental,zhang2018proof,An:19,ding2020280,PhysRevX.11.011038,roehsner2021probabilistic,roehsner2018quantum,pelet2022unconditionally}.

Nonetheless, all these schemes still have several limitations. Protocols utilizing orthogonal encoding require additional symmetrization steps which results in extra secure channels~\cite{Amiri:Secure:2016}. Therefore, to guarantee information-theoretic security, quantum key distribution (QKD) and one-time pad encryption are required between two receivers in orthogonal-type protocols~\cite{PhysRevA.95.042338,roberts2017experimental}.
Single-bit QDS schemes based on non-orthogonal encoding~\cite{Yin:practical:2016,lu2021efficient,Weng:21} are independent of additional QKD channels, but the signature rate is sensitive to the misalignment error of the quantum channel.
In addition, all these schemes can sign only a one-bit message in each round.
If one wants to sign a multi-bit message using single-bit QDS schemes, he needs to encode it into a new message string and sign the new string bit by bit~\cite{wang2015security,wang2017postprocessing,zhang2019high,cai2019cryptanalysis,yao2019multi,PhysRevA.103.012410}.
However, these solutions have not been completely proved as information-theoretically secure with the quantified failure probability, and the signature rate is very low and far from implementation for long messages with a lot of bits.

Recently, an efficient QDS scheme has been proposed based on secret sharing, one-time pad, and one-time universal hashing (OTUH)~\cite{yin2023experimental}.
Different from single-bit QDS protocols that require a long key string to sign a one-bit message, this OUTH-QDS protocol offers a method to directly sign the hash value of multi-bit messages through one key string with information-theoretic security, and thus drastically improves the QDS efficiency.
However, this framework requires perfect keys with complete secrecy, which is an expensive resource guaranteed by the complete procedure of QKD or quantum secret sharing (QSS). Accordingly, privacy amplification steps are required, thereby adding to the complexity of the algorithm and causing unendurable delays. 

Here, we point out that quantum keys with imperfect secrecy are  adequate for protecting the authenticity and integrity of messages in such a digital signature scheme. Accordingly, we propose a new OTUH-QDS protocol with imperfectly secret keys, utilizing only  asymmetric quantum keys without perfect secrecy to sign multi-bit messages. 
We demonstrate that our proposed scheme provides information-theoretic security for digital signature tasks and simulate the performance of our protocol.
The result reveals that our protocol outperforms other QDS schemes in terms of signature rate and transmission distance.  In a practical case of signing a megabit message, the proposed scheme has a \textcolor{black}{higher} signature rate of nearly eight orders of magnitude, compared with single-bit QDS schemes due to its robustness against message size. Moreover, we show that our scheme can significantly reduce the computational costs and delays of postprocessing owing to the removal of privacy amplification. %Our protocol is also robust against  and finite-size effect.
Furthermore, the proposed scheme is a general framework that can be applied to all existing QKD protocols. When utilizing the idea of two-photon twin-field QKD~\cite{Xie2023Scalable}, one of the most efficient QKD protocols, to execute our work, a transmission distance of 650 km can be achieved with a signature rate of 0.01 times per second.

To date, almost all quantum communication protocols such as QKD~\cite{bennett2014quantum,lo2012measurement,braunstein2012side,zhou2016making,lucamarini2018overcoming,ma2018phase,wang2018twin,liu2021homodyne,xie2022breaking,zeng2022mode,gu2022experimental}, QSS~\cite{fu2015long,gu2021differential,shen2023experimental}, and quantum conference key agreement~\cite{fu2015long,li2021finite} aim at generating quantum states among the parties and extract keys with perfect secrecy through complex postprocessing steps. 
Thereafter, these keys are then used to finish the corresponding cryptographic tasks such as private communication, secret sharing, and group encryption. 
In contrast, the proposed protocol offers a new approach to digital signature tasks that only require keys with imperfect secrecy through quantum optical communication. The troublesome postprocessing steps are thus moved out without relaxing the security assumption.  
This is the first instance of applying this kind of keys to cryptographic tasks with information-theoretic security.
We believe that our proposed solution can provide a feasible approach to the practical application of QDS and enlighten other applications of quantum keys with imperfect secrecy in a future quantum communication network.

The remainder of this paper is organized as follows. In Sec.~\ref{2} we review  OTUH-QDS scheme and introduce the motivation of this work. In Sec.~\ref{3} we propose our protocol with two approaches of universal hashing. In Sec.~\ref{SA} we give the security proof of authentication based on quantum keys with imperfect secrecy and then, the security analysis of the proposed QDS protocol. In Sec.~\ref{discussion} we discuss the performance of the proposed scheme and compare it with both single-bit QDS and OTUH-QDS schemes. Finally, we conclude the paper in Sec.~\ref{conclusion}.

\section{PRELIMINARIES}\label{2}

\begin{figure*}[t]
	\centering
	\includegraphics[width=0.8\linewidth]{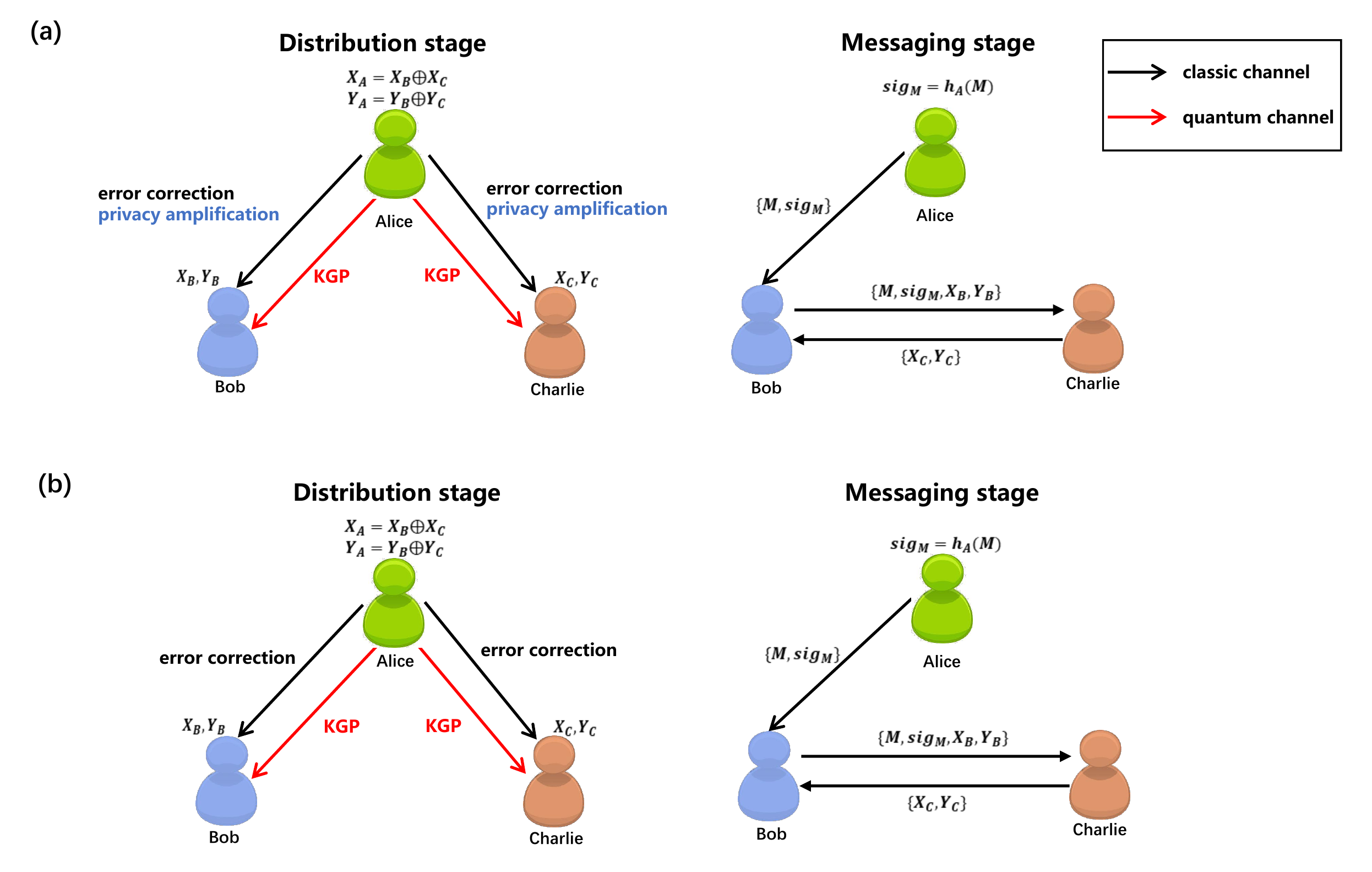}
		\caption{
		(a) OTUH-QDS~\cite{yin2023experimental}. In the distribution stage, Alice, Bob and Charlie share key bit strings with perfect secret sharing relationship through key generation protocol (KGP), error correction and privacy amplification. In the messaging stage Alice generates the signature through AXU hashing, and sends the message and signature to Bob. Bob then sends his keys and received information to Charlie, who will later send his keys to Bob. Ultimately, Bob and Charlie use their own and received keys to infer Alice keys and then perform AXU hashing to verify the signature.
	    (b) Schematic of the proposed protocol. In the distribution stage, the users only perform KGP and error correction to share keys with full correctness but some secrecy leakage. Their keys still hold secret sharing relationship. In the messaging stage the manipulation of classic information is analogous to that in OUTH-QDS.
        }
    \label{f1}
\end{figure*}

\subsection{OTUH-QDS protocol}

The schematic of OTUH-QDS~\cite{yin2023experimental} isreviewed herein. The protocol can be segmented into the distribution stage and messaging stage, consistent with single-bit QDS introduced in Appendix \ref{single-bit QDS}. The length of message is denoted as $m$.
The schematic of OTUH-QDS is shown in Fig.~\ref{f1}(a).

\subsubsection{distribution stage}

Alice, Bob, and Charlie each have two key bit strings $\{X_{a},~X_{b},~X_{c}\}$ with $n$ bits and $\{Y_{a},~Y_{b},~Y_{c}\}$ with $2n$ bits, satisfying the perfect correlation $X_{a}=X_{b}\oplus X_{c}$ and $Y_{a}=Y_{b}\oplus Y_{c}$, respectively. The distribution stage can be realized using quantum communication protocols, such as QKD and QSS.
It need to be mentioned that single-bit QDS requires only the quantum part of QKD protocols, also refered as key generation protocol (KGP).
In OTUH-QDS, the users need to perform additional error correction and privacy amplification steps after KGP.

\subsubsection{messaging stage}

(i) \emph{Signing of Alice}---First, Alice uses a local quantum random number, characterized by an $n$-bit string $p_a$, to randomly generate an irreducible polynomial $p(x)$ of degree $n$~\cite{menezes2018handbook}. Second, she uses the initial vector (key bit string $X_{a}$) and irreducible polynomial (quantum random number $p_{a}$) to generate a random linear feedback shift register-based (LFSR-based) Toeplitz matrix~\cite{krawczyk1994lfsr} $H_{nm}$, with $n$ rows and $m$ columns. Third, she uses a hash operation with $Hash$= $H_{nm} \cdot Doc$ to acquire an $n$-bit hash value of the $m$-bit document. Fourth, she exploits the hash value and the irreducible polynomial to constitute the $2n$-bit digest $Dig=(Hash||p_a)$. Fifth, she encrypts the digest with her key bit string $Y_{a}$ to obtain the $2n$-bit signature $Sig=Dig\oplus Y_{a}$ using OTP. Finally, she uses the public channel to send the signature and document $\{Sig,~Doc\}$ to Bob.
\textcolor{black}{Note that $Sig$ includes the information of the irreducible polynomial chosen for the hashing.}

(ii) \emph{Verification of Bob}---Bob uses the authentication classical channel to transmit the received $\{Sig,~Doc\}$, as well as his key bit strings $\{X_{b},~Y_{b}\}$, to Charlie. Thereafter, Charlie uses the same authentication channel to forward his key bit strings $\{X_{c},~Y_{c}\}$ to Bob. Bob obtains two new key bit strings $\{K_{X_{b}}=X_{b}\oplus X_{c},~K_{Y_{b}}=Y_{b}\oplus Y_{c}\}$ by the XOR operation. Bob exploits $K_{Y_{b}}$ to obtain an expected digest and bit string $p_b$ via XOR decryption. Bob utilizes the initial vector $K_{X_{b}}$ and irreducible polynomial $p_b$ to establish an LFSR-based Toeplitz matrix. He uses a hash operation to acquire an $n$-bit hash value and then constitutes a $2n$-bit actual digest. Bob will accept the signature if the actual digest is equal to the expected digest. Then, he informs Charlie of the result. Otherwise, Bob rejects the signature and announces to abort the protocol.

(iii) \emph{Verification of Charlie}---If Bob announces that he accepts the signature, Charlie then uses his original key along with the key sent to Bob to create two new key bit strings $\{K_{X_{c}}=X_{b}\oplus X_{c},~K_{Y_{c}}=Y_{b}\oplus Y_{c}\}$. Charlie employs $K_{Y_{c}}$ to acquire an expected digest and bit string $p_c$ via XOR decryption. Charlie uses a hash operation to obtain an $n$-bit hash value and then constitutes a $2n$-bit actual digest, where the hash function is an LFSR-based Toeplitz matrix generated by initial vector $K_{X_{c}}$ and irreducible polynomial $p_c$. Charlie accepts the signature only if the two digests are identical; otherwise, Charlie rejects the signature. 

\bigskip

The core point of this protocol is to realize the perfect bits correlation of the three parties, construct a completely asymmetric key relationship for them, and perform one-time almost XOR universal$_{2}$ (AXU) hashing, specifically, LFSR-based Toeplitz hashing, to generate the signature.
AXU hash functions is a special class of hash functions that can map an input value of arbitrary length into an almost random hash value with a preset length~\cite{carter1977universal}. 
The signature generated in OTUH-QDS is simply the AXU hash value of the long message to be signed, where the AXU hash function is determined by using only one string of Alice keys. 
After the distribution stage, Alice's, Bob's and Charlie's keys are completely secret and correct with the relationship of secret sharing. Bob (Charlie) can only obtain Alice's keys after he receives keys of Charlie (Bob).
Thus Bob can obtain no information of Alice's keys which decides the AXU hash function before transfering the message and signature to Charlie. Accordingly, Bob's forging attack under this protocol is equivalent to that against an authentication scenario where Alice sends an authenticated message to Charlie. It has been proved that such a message authentication scheme based on AXU hashing is information-theoretically secure~\cite{krawczyk1994lfsr}.
Consequently, forging attack is protected by one-time AXU hash functions and key relationship among three parties. 
From the perspective of Alice, Bob and Charlie's keys are totally symmetric when they verify the signature. Thus, Alice's repudiation attack is prevented as well.

\subsection{Motivation of this work}

Different from all single-bit QDS protocols that require a long key string to sign a one-bit message, OUTH-QDS offers a method to sign multi-bit messages through one key string with information-theoretic security, and thereby drastically improves the QDS efficiency.
Essentially, this advantage is introduced by AXU hash functions, which has been proved to be information-theoretically secure only under perfectly secret keys in previous studies. 
Thus, compared with single-bit QDS, OTUH-QDS requires extra error correction and privacy amplification steps to realize the perfect bits correlation in the distribution stage. 
These postprocessing steps especially  privacy amplification involves multiplication calculations on matrices with comparable length of data size, which introduces heavy computational costs and unpleasant delays in practical scenarios. For large-size data, the delays will become unendurable and constrain the practicality.

The process of AXU hashing is equivalent to the scenario where the input value decides the function, mapping the initial input keys into almost random output hash values. We notice that partial secrecy leakage of input value (keys) will be concealed in AXU hash value because of its randomness.
Thus, these imperfect keys with partial secrecy leakage will not undermine the authenticity of messages in a QDS scheme like OTUH-QDS.
Moreover, the integrity of messages is also not compromised.
Based on this concept, in this paper we propose a solution for OTUH-QDS protocols with imperfectly secret keys. In other words, we implement QDS with quantum keys without privacy amplification.
As the additional computational cost and delays of OTUH-QDS are primarily  introduced by privacy amplification, this concept can effectively reduce the weaknesses of OTUH-QDS and lay a ground for the future implementation of QDS in a quantum network.

The schematic of the proposed protocol is illustrated in Fig.~\ref{f1}(b).
In the distribution stage users only perform the error correction step after KGP, ensuring that their keys have no mismatches, and build a secret sharing relationship through Alice's XOR operation. 
The final keys will be randomly divided into several $n$-bit groups for AXU hashing. 
Each of these groups of keys contains full correctness and some secrecy leakage with an upper bound which can be estimated through finite-size analysis using experimental data in KGP. 
In the messaging stage, the rules of information exchange are consistent with that in OTUH-QDS. 
We will prove that the bit stings generated in our distribution stage are sufficient for AXU hashing and quantify the security bound in Sec.~\ref{SA}.
In addition, we give two solutions based on different types of AXU hash functions.

\begin{figure*}
	\centering
	\includegraphics[width=12.9cm]{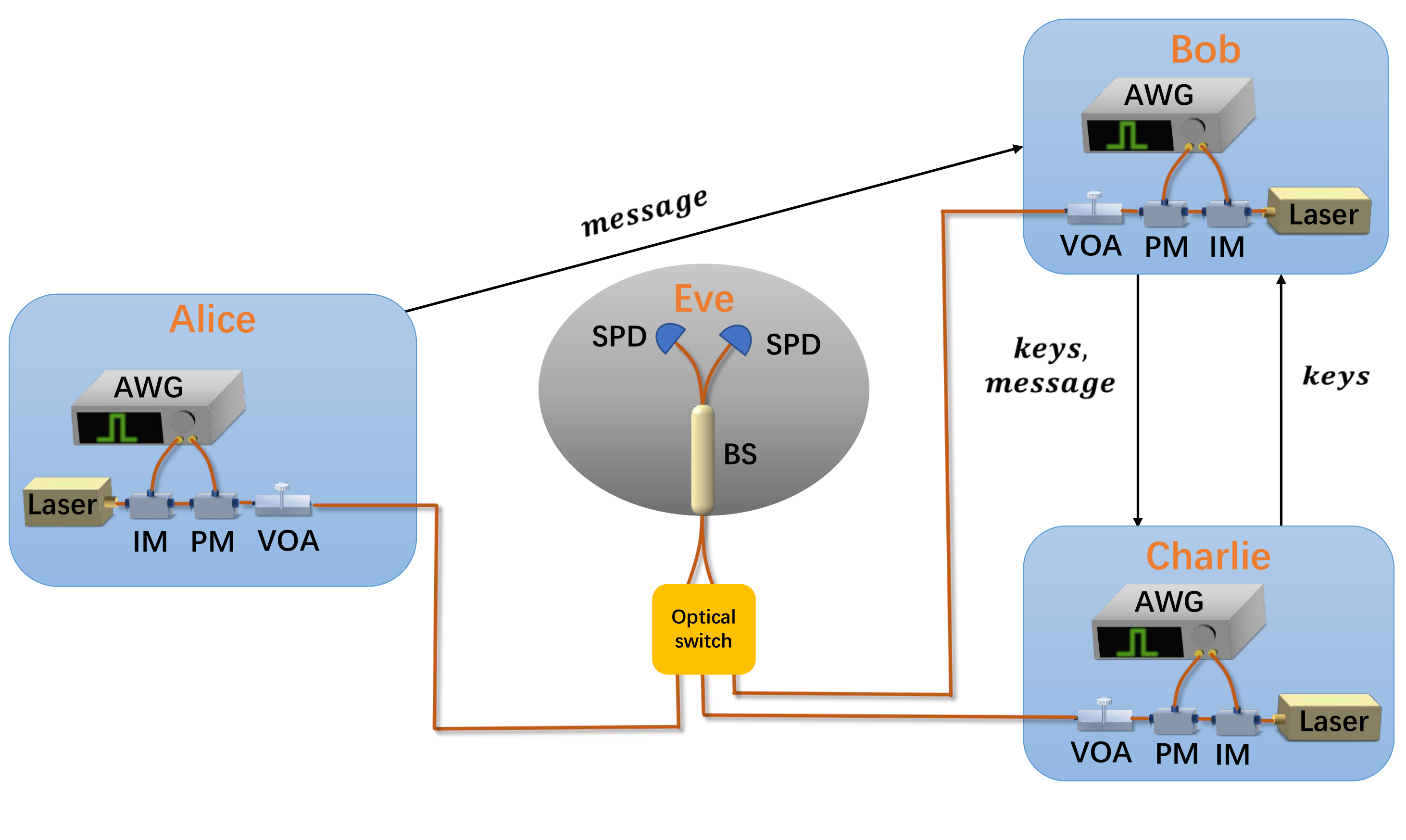}
	\caption{Schematic of the setup of the proposed QDS protocol. The red line represents quantum optical channel in the distribution stage, and the black arrow line represents the information exchange through classic authenticated channel in the messaging stage.
		(i) In the distribution stage, Alice, Bob and Charlie utilize a narrow-linewidth continuous-wave laser, intensity modulator (IM), phase modulator (PM), arbitrary wave generator (AWG) and variable optical attenuator (VOA) to prepare a phase-randomized weak coherent source with different intensities and phases. The signals from Bob and Charlie will both go through an optical switch. An untrusted relay Eve performs interference measurement on the signals from Alice and an optical switch with a beam splitter (BS) and a single-photon detector (SPD). After sifting, parameter estimation and error correction, Alice can share bit strings with Bob and Charlie, respectively.
		(ii) In the messaging stage, Alice transmits the desired message to Bob. Bob sends the message along with his keys to Charlie. Charlie will then send his keys to Bob. Then Bob verifies the signature by his own and received keys. If he accepts the signature, he will inform Charlie who will also verify the signature by his own and received keys. The signature is successfully validated if both Bob and Charlie accept it. }
	\label{setup}
\end{figure*}

\section{QDS PROTOCOL}\label{3}

A schematic of setups of the proposed QDS protocol is illustrated herein and illustrated in Fig.~\ref{setup}.

\subsection{Distribution stage}

Our proposal is a general framework in which KGP can be derived from any type of QKD protocol. As an example, the proposed scheme is demonstrated based on two-photon twin-field (TP-TF) QKD~\cite{Xie2023Scalable}.
In the distribution stage, Alice–-Bob and Alice–-Charlie independently implement TP-TF KGP (TP-TFKGP for simplicity) to share key bit strings.
We remark that in this three-party protocol the process of Alice–-Bob and Alice–-Charlie are independent, and can be performed separately. The difficulty of the experiment is the same as two-party QKD protocols. Specifically, TP-TFKGP utilizes the idea of two-photon interference to distribute quantum states. Consequently, the performance is independent of probability and intensity for each user, meanwhile having high misalignment error tolerance. The protocol is thus unaffected by the addition or deletion of users \textcolor{black}{(as long as the number of users is on less than three)}, highly versatile and suitable for future quantum metropolitan networks.

{\it{1.~Preparation.}}
At each time bin $i\in\{1,2,\ldots,N\}$, Alice and Bob (Alice and Charlie) each independently prepare a weak coherent pulse $\ket{e^{\textbf{i}(\theta_{x}^{i}+r_{x}^{i}\pi)}\sqrt{k_{x}^i}}$ with probability $p_{k_x}$, where the subscript $x\in\{a, b, c\}$ represents the user Alice, Bob or Charlie, the phase $\theta_{x}^{i}$  $\in[0,2\pi)$, classical bit $r_{x}^{i}\in\{0, 1\}$, \textcolor{black}{intensity $k_{x}^i\in \{\mu_{x},~\nu_{x},~\mathbf{o}_{x},~\hat{\mathbf{o}}_{x}\}$ (represent signal, decoy, preserve-vacuum and declare-vacuum intensity, $\mu_{x}>\nu_{x}>\mathbf{o}_{x} =\hat{\mathbf{o}}_{x} = 0$) are chosen randomly.} 
Then Alice and Bob (Alice and Charlie) transmit the corresponding pulses to the untrusted relay Eve through insecure quantum channels, respectively.
In addition, they send a bright reference light to Eve to measure the phase noise difference $\phi_{ab}^i$ ($\phi_{ac}^i$).

{\it{2.~Measurement.}}
Eve performs interference measurements on every received pulse pair with a beam splitter and two detectors.
If one and only one detector clicks, Eve announces that she obtained a successful detection event and which detector clicked.
\textcolor{black}{In the following we use the brace with the information of users' intensity selection in it to distinguish these events. For example, $\{\mu_a,\mathbf{o}_{b}\}$ represents the events that Alice selects signal intensity and Bob selects vacuum intensity.}

{\it{3.~Sifting.}}
\textcolor{black}{Here we only list the sifting process between Alice and Bob for simplicity since Alice--Bob and Alice--Charlie are symmetric. Alice and Charlie will sift their successful detection events following the same approach.}

\textcolor{black}{All successful events are segmented into two parts. The first part is those when neither Alice nor Bob selects the decoy or declare-vacuum intensity,
i.e., $\{\mu_a,\mathbf{o}_{b}\}$, $\{\mu_a,\mu_b\}$, $\{\mathbf{o}_{a},\mu_b\}$, and $\{\mathbf{o}_{a},\mathbf{o}_{b}\}$, which will be used for generating data in the $Z$ basis to form the key. The other successful events, i.e., the second part, are used for  estimating parameters.
For the first part of events,} Alice randomly matches a time bin $i$ of intensity  $\mu_a$  with another time bin $j$ of intensity $\mathbf{o}_{a}$. Thereafter she sets her bit value  as 0 (1) if
$i< j~(i>j)$, and informs the serial numbers $i$ and $j$ to Bob. In the corresponding time bins, if Bob chooses intensities $k_b^{\min\{i,j\}}=\mu_b$~$(\mathbf{o}_{b})$, and $k_b^{\max\{i,j\}}=\mathbf{o}_{b}$~$(\mu_b)$, he sets his bit value as 0 (1). Bob announces to abort the event where $k_b^i$ $=$ $k_b^j=\mathbf{o}_{b}$ or $\mu_b$. To conclude, the preserved events in the $Z$ basis are sifted as $\{\mu_a \mathbf{o}_{a},~ \mathbf{o}_{b}\mu_b\}$, $\{\mu_a\mathbf{o}_{a},~\mu_b\mathbf{o}_{b}\}$, $\{\mathbf{o}_{a}\mu_a,~ \mathbf{o}_{b}\mu_b\}$, and $\{\mathbf{o}_{a}\mu_a,~\mu_b\mathbf{o}_{b}\}$.

\textcolor{black}{For the second part of events, Alice and Bob communicate their intensities and phase information with each other via
an authenticated channel.}
Define the global phase difference at time bin $i$ as $\theta^i:= \theta_a^i- \theta_b^i +\phi_{ab}^i$. Alice and Bob keep detection events $\{\nu_a^i,~\nu_b^i\}$ only if $\theta^i~\in[-\delta,\delta]\cup[\pi-\delta,\pi+\delta]$. They randomly select two retained detection events that satisfy $\left|\theta^i -\theta^j\right|= 0 $ or $\pi$, and then match these two events , denoted as $\{\nu_a^i\nu_a^j,\nu_b^i\nu_b^j\}$. By calculating classical bits $r_a^i \oplus r_a^j $ and $r_b^i \oplus r_b^j $, Alice and Bob extract a bit value in the $X$ basis, respectively.
Subsequently,  Bob always flips his bit in the $Z$ basis.
In the $X$ basis, Bob flips part of his bits to correctly correlate them with those of Alice.
To be specific, when the global phase difference between two matching time bins is 0 ($\pi$) and the two clicking detectors announced by Eve are different (same), Bob will flip his bits. Otherwise, Bob will directly save his bits for later use.
\textcolor{black}{The other events in the second part is used for decoy analysis.}
%Similarly, Alice and Charlie will sift their successful detection events following the same approach.

{\it{4.~Parameter estimation.}} Alice and Bob (Alice and Charlie) form the $n_Z$-length raw key bit from the random bits under the $Z$ basis. The remaining bits in the  $Z$ basis are used to estimate the bit error rate  $E^z$. Further, they communicate all bit values in the $X$ basis to obtain the total number of errors. The decoy-state method~\cite{wang2005beating,lo2005decoy} is used to estimate the number of vacuum events in the $Z$ basis $s_{0\mu_b}^{z}$, the count of single-photon pairs  $s_{11}^z$, and the phase error rate of single-photon pairs $\phi_{11}^z$ on the $Z$ basis.

{\it{5.~Error correction and examination.}} Alice and Bob (Alice and Charlie) distill final keys by utilizing an error correction algorithm with $\varepsilon_{\rm{cor}}$-correctness.~\cite{brassard1993secret,yan2008information}
The size of the distilled key remains $n_Z$, and the unknown information of a possible attacker can be expressed as $\mathcal{H}$.
Alice then randomly disturbs the orders of the distilled key and publicizes the new order to Bob (Charlie) through the authenticated channel.
Subsequently, Alice and Bob (Alice and Charlie) divide the final keys into several $n$-bit strings, each of which is used to perform a task in the messaging stage. The grouping process can be considered as a random sampling. More details are shown in Sec. \ref{SA1} and Appendix \ref{A31}.
%This ensures that the attacker cannot know the position of each bit of one concrete group beforehand, and thus can not perform a

\subsection{Messaging stage}
Various AXU hash functions can be employed in the messaging stage of the proposed protocol by following the framework presented in Fig.~\ref{f1}(b).
To demonstrate the detailed procedure, we here present two specific approaches to the messaging stage utilizing LFSR based Toeplitz hashing and generalized division hashing, respectively.
LFSR-based Toeplitz hashing is highly compatible with the hardware systems whereas generalized division hashing is more suitable for realizing software systems. In a practical case we select either of methods of hashing depending on the different application environments of users.
The message to be signed is denoted  as $M$.
For each $M$ if using LFSR-based Toeplitz hashing Alice generates six bit strings $\mathbb{X}_B,~\mathbb{X}_C,~\mathbb{Y}_B,~\mathbb{Y}_C,~\mathbb{Z}_B,~\mathbb{Z}_C$, each  of length n. If choosing generalized division hashing in the messaging stage, Alice will only generate four bit strings $\mathbb{X}_B,~\mathbb{X}_C,~\mathbb{Y}_B,~\mathbb{Y}_C$.
The subscripts represent the participants performing KGP with Alice, where B represents Bob and C represents Charlie.
Thereafter,  Alice will generate $\mathbb{X}_{a}=\mathbb{X}_{b}\oplus \mathbb{X}_{c}$, $\mathbb{Y}_{a}=\mathbb{Y}_{b}\oplus \mathbb{Y}_{c}$, and $\mathbb{Z}_{a}=\mathbb{Z}_{b}\oplus \mathbb{Z}_{c}$ as her own key strings.
For the scheme with LFSR-based Toeplitz hashing the signature rate is 
\be\label{eq1}
R_{\rm{LFSR}}=n_Z/3n,
\ee 
whereas for generalized division hashing there is 
\be\label{eq2}
R_{\rm{GDH}}=n_Z/2n.
\ee

\subsubsection{Utilizing LFSR-based Toeplitz hashing}

\begin{definition}\label{def1}
	LFSR-based Toeplitz hash functions:
	LFSR-based Toeplitz hash functions can be expressed as $h_{p,s}(M)=H_{nm}  M$, where $p,s$ determines the function and $M=(M_0,M_1,...,M_{m-1})^T$ is the message in the form of an $m$-bit vector.
	The process of generating LFSR-based Toeplitz hash function is detailed as follows.
	
	A randomly selected irreducible polynomial of order $n$ in the field GF(2), $p(x)$, determines the construction of LFSR. $p(x)=x^n+p_{n-1}x^{n-1}+...+p_1x+p_0$ can be characterized by its coefficients of order from $0$ to $n-1$, i.e., $p=(p_{n-1},p_{n-2},...,p_1,p_0)$. For the initial state $s$ which is also represented as an n-bit vector $s=(a_n,a_{n-1},...,a_2,a_1)^T$, the LFSR will be performed n times to generate n vectors. Specifically, it will shift down every element in the previous column, and add a new element to the top of the column. For instance, the LFSR transforms $s$ into $s_1=(a_{n+1},a_n,...,a_3,a_2)^T$, where $a_{n+1}=p \cdot s$, and likewise, transforms $s_1$ to $s_2$. Then the m vectors $s,s_1,...,s_{m-1}$ will together construct the Toeplitz matrix $H_{nm}=(s,s_1,...,s_{m-1})$, and the hash value of the message is $H_{nm}  M$.
\end{definition}

(i)  Alice obtains a string of random numbers through a quantum random number generator and uses it to randomly generate a monic irreducible polynomial in GF(2) of order $n$, denoted as $p(x)$. $p(x)$ can be characterized by its coefficients of order from $0$ to $n-1$, i.e., an n-bit string, denoted by $p_a$. Details of generating $p(x)$ can be found in Appendix~\ref{A2}.

(ii)
Alice uses her key bit string $\mathbb{Y}_a$ and $p(x)$ to perform LFSR-based Toeplitz hashing and generates an n-bit hash value $Dig=H_{\mathbb{Y},p_a}(M)$, and encrypts it by $\mathbb{Z}_a$ to obtain the final signature $Sig=Dig\oplus \mathbb{Z}_a$. In addition, Alice encrypts $p_a$ by the key set $\mathbb{X}_a$ to obtain the encrypted string $p=p_a \oplus \mathbb{X}_a$.
\textcolor{black}{Here we adopt a different expression from that in OUTH-QDS that we independently list the hash value as $Dig$ and the coefficients of the irreducible polynomial as $p_a$, i.e., $Sig$ does not include the information of the irreducible polynomial to avoid misunderstanding.}
She then transmits $\{Sig,~p,~M\}$ to Bob through an authenticated classical channel.

(iii) Bob transmits $\{Sig,~p,~M\}$ as well as his key bit strings $\{\mathbb{X}_{b},~\mathbb{Y}_{b},~\mathbb{Z}_b\}$ to Charlie so as to inform Charlie that he has received the signature. Thereafter, Charlie forwards his key bit strings $\{\mathbb{X}_{c},~\mathbb{Y}_{c},~\mathbb{Z}_c\}$ to Bob. These data are all transmitted through an authenticated channel. Bob obtains three new key bit strings $K_{\mathbb{X}_{b}}=\mathbb{X}_{b}\oplus \mathbb{X}_{c},~K_{\mathbb{Y}_{b}}=\mathbb{Y}_{b}\oplus \mathbb{Y}_{c}$, and $K_{\mathbb{Z}_{b}}=\mathbb{Z}_{b}\oplus \mathbb{Z}_{c}$ using the XOR operation. He exploits $K_{\mathbb{X}_{b}}$ and $K_{\mathbb{Z}_{b}}$ to obtain the expected digest and string $p_b$ via XOR decryption. He utilizes $K_{\mathbb{Y}_{b}}$  and $p_b$ to establish an LFSR-based Toeplitz matrix and derive an actual digest via a hash operation. Bob will accept the signature if the actual digest is equal to the expected digest. Then he informs Charlie of the result.

(iv) If Bob announces that he accepts the signature, Charlie creates three new key bit strings $K_{\mathbb{X}_{c}}=\mathbb{X}_{b}\oplus \mathbb{X}_{c},,~K_{\mathbb{Y}_{c}}=\mathbb{Y}_{b}\oplus \mathbb{Y}_{c}$ and $K_{\mathbb{Z}_{c}}=\mathbb{Z}_{b}\oplus \mathbb{Z}_{c}$ using his original key and that received from Bob. He employs $K_{\mathbb{X}_{c}}$ and $K_{\mathbb{Z}_{c}}$ to acquire the expected digest and variable $p_c$  via XOR decryption. Charlie obtains an actual digest via hash operation, where the hash function is an LFSR-based Toeplitz matrix generated by $K_{\mathbb{Y}_{c}}$ and $p_c$. Charlie accepts the signature if the two digests are identical.

\subsubsection{Utilizing generalized division hashing}

\begin{definition}
	Generalized division hash functions:
	The generalized division hash functions can be expressed as $h_{P}(M)=M(x) \cdot x^{n/k}$ mod $P(x)$, where $P(x)$ is a monic irreducible polynomial of order $n/k$ in the field GF($2^k$), $M$ is the message and $M(x)$ is  the  polynomial of order $m/k$ in GF($2^k$) with every coefficient corresponding to $k$ bits of $M$. The calculation is also performed in GF($2^k$). The final result is a polynomial of order $n/k$ in  field GF($2^k$), and can be transformed into an $n$-bit strings.~\cite{shoup1996fast}
	
\end{definition}
Commonly, $k$ is set as $k=2^x$ for simplicity, where $x$ is a positive integer.
In the current scheme, we select $k=2^3=8$.

(i) In this case, Alice selects  $\mathbb{X}_{a}=\mathbb{X}_{b}\oplus \mathbb{X}_{c}$ and $\mathbb{Y}_{a}=\mathbb{Y}_{b}\oplus \mathbb{Y}_{c}$  as her own key sets.
Alice first obtains a string of random numbers through a quantum random number generator and uses it to randomly generate a monic irreducible polynomial in GF($2^8$) of order $n/8$, denoted by $P(x)$. The generation process of $p(x)$ are detailed in Appendix~\ref{A2}. $P(x)$ can be characterized by its coefficients of order from 0 to $n/8-1$. By encoding each coefficient into an 8-bit string, we can use an n-bit string to express $P(x)$, denoted as $P_a$.
Subsequently, Alice encrypts $P_a$ by the key set $\mathbb{X}_a$ to obtain the encrypted string $P=P_a \oplus \mathbb{X}_a$

(ii)
Alice uses $P(x)$ to perform the generalized division hashing~\cite{shoup1996fast} to obtain an $n$-bit hash value $Dig=h_{P_a}(M)$.
She encrypts $Dig$ by $\mathbb{Y}_a$ to derive the signature $Sig=Dig \oplus \mathbb{Y}_a$ and transmits the message along with the obtained signature $\{Sig,~p,~M\}$ to Bob.

step (iii) and (iv) are similar to those utilizing LFSR-based Toeplitz hashing. Bob and Charlie will exchange their key strings in turn  through an authenticated channel and examine their expected and received digests.

\bigskip
This summarizes the entire procedure of the proposed protocol.
Note that the TP-TFKGP can be replaced by any other KGP such as BB84-KGP or sending-or-not-sending (SNS)-KGP. Actually, in the distribution stage Alice shares bit strings with Bob and Charlie in the relationship of secret sharing.
Thus, the distribution stage can also be performed based on QSS without employing the privacy amplification step.

\section{Security analysis}
\label{SA}

Similar to OTUH-QDS, the core point of the proposed protocol is the security of the authentication based on AXU hashing, which directly protects the security of QDS against fogery~\cite{yin2023experimental}. However, the security of our protocol differs because of the information leakage during the distribution stage. In this section we first analyze the  success probability of an attacker guessing a key string generated in the distribution stage, and thereafter provide a more detailed security analysis of AXU hashing under imperfect keys with partial secrecy leakage, and finally demonstrate the security of our protocol.

\subsection{Guessing probability of the attacker}\label{SA1}
Unlike QKD that generates keys with perfect secrecy, in our protocol the keys are imperfectly secret. Any possible attackers may obtain partial information on the keys. After the distribution stage, users share keys in the form of several $n$-bit strings.
We need to quantify the information leakage and bound the maximum probability of the attacker guessing such a string of keys.
Suppose an n-bit key string as $\mathbb{X}$ and the attacker's system is $\mathbb{B}$.
We consider a general attack scenario where attackers can execute any  entangling operations on the system of any or all states, obtain a system $\rho_B^x$ and perform any positive operator-valued measure $\{E_{B}^{x}\}_x$ on it. 
The probability that the attacker correctly guesses $\mathbb{X}$  using an optimal strategy is denoted as $P_{\rm{guess}}(\mathbb{X}|\mathbb{B})$. According to the definition of min-entropy in Ref.~\cite{konig2009operational},
\begin{equation}
	P_{\rm{guess}}(\mathbb{X}|\mathbb{B})=\max_{\{E_{B}^{x}\}_x} \sum_{x}P_x  \operatorname{tr}(E_{B}^{x} \rho_{B}^{x}) =  2^{-H_{\rm{min}}(\mathbb{X}|\mathbb{B})_\rho},
\end{equation}
where $H_{\rm{min}}(\mathbb{X}|\mathbb{B})_\rho$ is the min-entropy of $\mathbb{X}$ and $\mathbb{B}$. If $\mathbb{X}$ is generated in the distribution stage of our protocol, $H_{\rm{min}}(\mathbb{X}|\mathbb{B})_\rho$ can be estimated by 
\be
H_{\rm{min}}(\mathbb{X}|\mathbb{B})_\rho=\mathcal{H}_n.
\ee
Thus, we have the relationship
\begin{equation}
	P_{\rm{guess}}(\mathbb{X}|\mathbb{B})=2^{-\mathcal{H}_n},
\end{equation}
which means that the attacker can correctly guess $\mathbb{X}$ with a probability no more than $2^{-\mathcal{H}_n}$. $\mathcal{H}_n$ is the total unknown information of the $n$-bit string and can be upper bounded by parameters estimated in the distribution stage
\begin{equation}
	\mathcal{H}_n\le \underline{s}_{0\mu_b}^{zn}+\underline{s}_{11}^{zn}\left[1-H(\underline{\phi}_{11}^{zn})\right]-n f H(E^z),
	\label{1}
\end{equation}
where $f$ is the error correction efficiency, $\underline{s}_{0\mu_b}^{zn}$ and $\underline{s}_{11}^{zn}$ are the lower bounds of vacuum events and single-photon pairs events in the $n$-bit string, respectively, and $\underline{\phi}_{11}^{zn}$ is the upper bound of  the phase error rate of single-photon pairs in the $n$-bit string. More details of calculation are shown in Appendix \ref{A31}.

\subsection{Security of authentication based on hashing}\label{SA2}
In our QDS schemes, hashing is used to perform the authentication task. Thus we first consider the authentication scenario where the sender generates a signature $Sig=h(M)\oplus r$ as message authentication code,  and sends $\{M,~Sig\}$ to the recipient. The attacker can intercept and capture $\{M,~Sig\}$, tamper a new message and signature $\{M',~Sig'\}$, and send it to the recipient, who will examine whether $Sig'=h(M')\oplus r$ before accepting it.
The attacker succeed \textcolor{black}{if and only if (iff) a combination} $\{m,~t\}$ is select with the relationship $h(m) = t$, and $\{M'=M\oplus m,~Sig'=Sig\oplus t\}$ is sent to the recipient. In this case, the recipient will accept the message because of the relationship $h(M\oplus m) = h(M)\oplus h(m)= Sig\oplus t$.
It should be mentioned that $m\ne 0$ due to the requirement for a valid forge.

Suppose keys generated in the distribution stage of our protocol, i.e., keys with partial information leakage, are used to perform this authentication task, and
define $\epsilon$ as the success probability of the attacker under this scenario.
We should consider three types of possible attacks. 
The first one is to randomly generate ${m,t}$. It is a trivial strategy whose success probability is only 
\be
\epsilon_1=2^{-n}.
\ee
The other two types of attacks are guessing keys that decide the hash function and recovering the function from signatures.

\subsubsection{Attack of guessing keys}

The LFSR-based Toeplitz hash function is represented as $h_{p,s}(M)$= $H_{nm} \cdot M$, where $H_{nm}$ is determined by the two bit strings $p$ and $s$~\cite{krawczyk1994lfsr}. Herein we follow the terminology in the messaging stage of the proposed protocol where $p$ is actually $p_a$ encrypted by $\mathbb{X}_a$, $s$ is $\mathbb{Y}_a$, and the hash value $Dig$ is encrypted by $\mathbb{Z}_a$. 
We show that guessing only $\mathbb{X}_a$ or in other words, guessing only $p_a$ is enough to execute an optimal attack by a proposition. 

\begin{proposition}\label{lem1}For the LFSR-based Toeplitz hash function $h_{p,s}(M)$= $H_{nm} \cdot M$, if $p(x) | M(x)= M_{m-1}x^{m-1}+...+M_1 x +M_0 $, then $h_{p,s}(M)=0$.
\end{proposition}

The proof of this proposition is shown in Appendix~\ref{A22}. It means that the attacker can easily generate a message $m$ satisfying the relationship $h(m)=0$ if he knows $p$.
In the scenario described above, suppose the attacker obtains a string $\mathbb{X}_g$ as his estimation of $\mathbb{X}_a$. He can decrypt it to obtain $p_g$ as his guessing of $p_a$ and transform $p_g$ into a polynomial $p_g(x)$.
Thereafter the attacker can easily generate a bit string $m$ satisfying $p_g(x) | m(x)$, and there is the relationship $h(m) = 0$ if $p_g=p_a$ (or equivalently $\mathbb{X}_g=\mathbb{X}_a$) according to Proposition \ref{lem1}. Then he can tamper the message into $M\oplus m$ without changing the signature. $\{M+m,Sig\}$ will pass the authentication test if $\mathbb{X}_g=\mathbb{X}_a$.
As $m(x)$ is m-order and the polynomial is n-order, the attacker can select no more than $m/n$ polynomials and multiply them to consist his choice of $m(x)$. In other words, he can guess the string $\mathbb{X}_a$ for no more than $m/n$ times.
It must be considered that the attacker knows $p_a$ is irreducible, so he will only choose those guesses that satisfy $p_a$ is irreducible.
The success probability of this optimized strategy can be expressed as 
\be
    P_1= \frac{m}{n}\cdot P(\mathbb{X}_a=\mathbb{X}_g|p_g\in \mathcal{I}),
\ee 
where $P(A | B)$ represents the probability of event $A$ under the condition that event $B$ occurs,
and $\mathcal{I}$ denotes the set of all irreducible polynomials of order n in GF(2). 
The cardinal number of $\mathcal{I}$, i.e., the number of all n-order irreducible polynomials in GF(2), is more than $2^{n-1}/n$. Thus $P(p_g\in \mathcal{I})\le (2^{n-1}/n)/2^n=1/2n$. It is obvious that $P(\mathbb{X}_a=\mathbb{X}_g,p_g\in \mathcal{I})=P(\mathbb{X}_a=\mathbb{X}_g)$ because if $\mathbb{X}_a=\mathbb{X}_g$ then $p_g=p_a\in \mathcal{I}$. 
Then we can obtain the upper bound of the success probability of this type of attack, denoted as $\epsilon_{\rm{LFSR}}$, 
\begin{align*}
   P_1=& \frac{m}{n}\cdot \frac{P(\mathbb{X}_a=\mathbb{X}_g)}{P(p_g\in \mathcal{I})} \\
   \le & \frac{m}{n}\cdot \frac{2^{-\mathcal{H}_n}}{\frac{1}{2n}} \\
   =& m\cdot 2^{1-\mathcal{H}_n}=\epsilon_{\rm{LFSR}}.
\end{align*} 
The attacker can also guess the strings $\mathbb{X}_a$ and $\mathbb{Y}_a$ to obtain $p$ and $s$ so that he can guess the hash function and make a successful attack for certainty. Under this circumstance his success probability is no more than $\epsilon_{\rm{LFSR}}$,
\begin{align*}
	P_2=& P(\mathbb{X}_a=\mathbb{X}_g,\mathbb{Y}_a=\mathbb{Y}_g|p_g\in \mathcal{I}) \\
	\le & P(\mathbb{X}_a=\mathbb{X}_g|p_g\in \mathcal{I}) \\
	\le& \epsilon_{\rm{LFSR}}.
\end{align*}  
 
The generalized division hash function $h_{P}(M)$= $m(x) \cdot x^{n/8}$ mod $P(x)$ is determined only by $P$. As earlier, we also follow the terminology in the proposed protocol that $P$ is  $P_a$ encrypted by $\mathbb{X}_a$ and the hash value $Dig$ is encrypted by $\mathbb{Y}_a$.
The attacker's strategy is to guess a string 
 $\mathbb{X}_g$ such that he can obtain $P_g$ and then forge a message. Analogous to the analysis discussed above, the upper bound of the success probability is defined as
\begin{align}
   \epsilon_{\rm{GDH}}=\frac{m}{n}\cdot \frac{2^{-\mathcal{H}_n}}{\frac{4}{n}}=m\cdot 2^{-2-\mathcal{H}_n}.
\end{align} 
The only difference in the calculation is that there are at least $2^{n-1}/(n/8)$ irreducible polynomials of order $n/8$ in GF($2^8$), so $P(P_g\in \mathcal{I})\ge (2^{n-1}/(n/8))/2^n=4/n$.  
%Finally we can obtain the failure probability 

\subsubsection{Attack of recovering keys from signature} 

The attacker can attempt to recover the desired keys from the captured signature. In both kinds of hashing the hash value is encrypted to generate the signature. Thus the attacker must first guess the corresponding key strings ($\mathbb{Z}_a$ in LFSR-based Toeplitz hashing or $\mathbb{Y}_a$ in generalized division hashing) and then perform the recovering algorithm.
The success probability of this strategy is no more than that only guessing the bit string ($P(\mathbb{Z}_a=\mathbb{Z}_g)$ or $P(\mathbb{Y}_a=\mathbb{Y}_g)$) and is obviously no more than $\epsilon_{\rm{LFSR}}$ or $\epsilon_{\rm{GDH}}$.
 
\bigskip 
 
In conclusion, the optimal strategy on LFSR-based Toeplitz hashing and  generalized division hashing (GDH) is to guess the key string that encrypts the polynomial. 
We can quantify the upper bound of failure probability of authentication based on both types of hashing with imperfect keys of secrecy leakage:
\begin{align}\label{eq10}
	\epsilon_{\rm{LFSR}}=&m\cdot 2^{1-\mathcal{H}_n},\\
	\epsilon_{\rm{GDH}}=&m\cdot 2^{-2-\mathcal{H}_n}\label{eq11}.
\end{align}

\subsection{Security of the QDS scheme}

Finally, we analyze the security in the QDS scheme which contains three parts, robustness, repudiation, and forgery.

\subsubsection{Robustness.}
The honest run abortion means the protocol is aborted when all parties are honest.
It occurs only when Alice and Bob (or Charlie) share different key bits after the distribution stage. In the proposed protocol Alice and Bob (Charlie) perform error correction in the distribution stage. Thus, they share the identical final key, and the honest run occurs only at the case where errors occur. The robustness bound is $\epsilon_{\rm{rob}}=2\epsilon_{\rm{cor}}+2\epsilon'$, where $\epsilon_{\rm{cor}}$ is the failure probability of the error correction protocol in the distribution stage, and $\epsilon'$ is the probability that error occurs in classical message transmission.
Remark that  we assume $\epsilon'=10^{-11}$ for simplicity since it is a parameter of classical communication.

\subsubsection{Repudiation.}
Alice successfully repudiates when Bob accepts the message while Charlie rejects it.
For Alice’s repudiation attacks, Bob and Charlie are both honest and symmetric and possess the same new key strings. They will converge on the same decision for the same message and signature. In other words, when Bob rejects (accepts) the message, Charlie also rejects (accepts) it. Repudiation attacks succeed only when errors occur in one of the key exchange steps.
Thus, the repudiation bound is $\epsilon_{\rm{rep}}=2\epsilon'$.

\subsubsection{Forgery.}
Bob forges successfully when Charlie accepts the tampered message forwarded by Bob. According to the proposed protocol, Charlie accepts the message iff Charlie obtains the same result through one-time pad decryption and one-time AXU hash functions. In principle, this is the same as an authentication scenario in Sec. \ref{SA2} where Bob is the attacker attempting to forge the information sent from Alice to Charlie.
Therefore, the probability of a successful forgery $\epsilon_{\rm{for}}$ can be determined by the failure probability of hashing, i.e., one chooses two distinct messages with identical hash values. For the scheme utilizing LFSR-based Toeplitz hash $\epsilon_{ \rm{for}}=m\cdot 2^{1-\mathcal{H}_n}$, and for generalized division hashing $\epsilon_{\rm{for}}=m\cdot 2^{-2-\mathcal{H}_n}$.

\bigskip
The total security bound of QDS, i.e., the maximum failure probability of the protocol, is $\epsilon= \max\{\epsilon_{\rm{rob}},\epsilon_{\rm{rep}},\epsilon_{\rm{for}}\}$.

\begin{table}[b]
	\centering
	\caption{Simulation parameters. $\eta_{d}$ and $p_{d}$ denote the detector efficiency and dark count rate, respectively. $e_d$ represents the misalignment error rate. $N$ is the data size. $\alpha$ is the attenuation coefficient of the fiber. $f$ is the error correction efficiency. $\epsilon$ is the failure probability of QDS schemes.}
	\begin{tabular}[b]{@{\extracolsep{15pt}}ccccccc}
		\hline
		\hline
		$\eta_{d}$  & $p_{d}$ & $e_d$  & $N$   & $\alpha$ & $f$  & $\epsilon$\\
		\hline%\xrowht{7pt}
		$70\%$ & $10^{-8}$ & $0.02$  & $10^{13}$   & $0.165$ & $1.1$  &
		$10^{-10}$\\
		\hline
		\hline
	\end{tabular}
    \label{tabel2}
\end{table}

\section{Discussion}\label{discussion}

\begin{figure}[t]
	\centering
	\includegraphics[width=8.6cm]{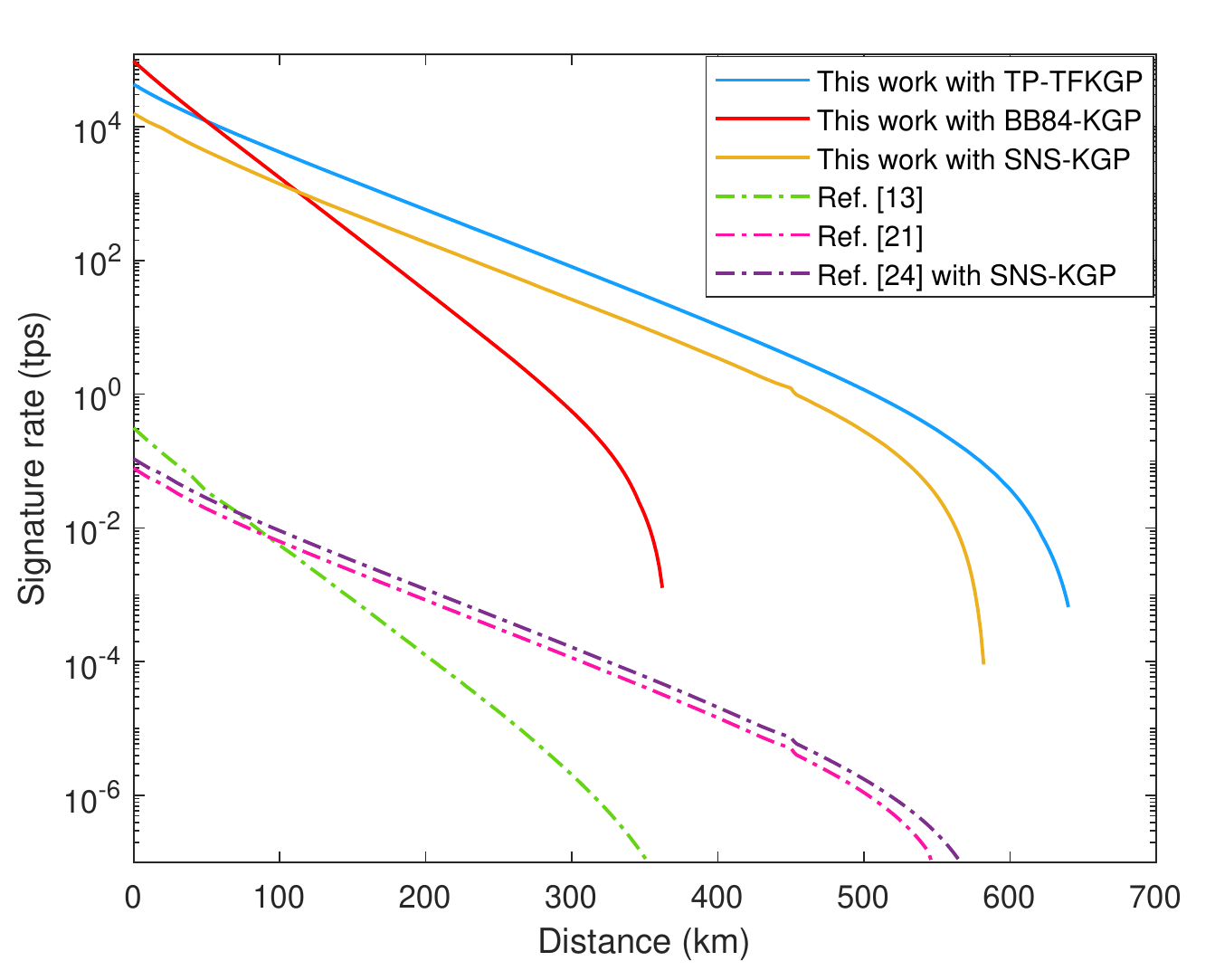}
	\caption{Signature rates of the proposed protocol with TP-TFKGP, BB84-KGP, SNS-KGP, and decoy state BB84-QDS~\cite{Amiri:Secure:2016}, SNS-QDS~\cite{zhang2021twin}, SNS-QDS with random pairing~\cite{qin2022quantum} with the message size of 1 Kb. In the proposed protocol we use generalized division hashing in messaging stage.
    The repetition rate of the laser is 1 GHz. The distances between Alice--Bob and Alice--Charlie are assumed to be the same. The data size N is $10^{13}$ and the security bound is $10^{-10}$.
	}
	\label{f3}
\end{figure}

From Eqs. \eqref{eq1}, \eqref{eq2}, \eqref{eq10}, and \eqref{eq11}, 
there are just differences in a constant $2/3$ between two signature rates and a constant 8 between two security parameters. The difference between the two approaches is trivial.
For simplicity, we only discuss the protocol with generalized division hashing in this section.

To demonstrate the \textcolor{black}{advantage} of the current proposal, we first build our protocol based on BB84-KGP, SNS-KGP and TP-TFKGP, and compare them with decoy-state BB84-QDS~\cite{Amiri:Secure:2016} and SNS-QDS~\cite{zhang2021twin} which are single-bit QDS protocols based on BB84-KGP and SNS-KGP. We also compare SNS-QDS with random pairing~\cite{qin2022quantum}, which improves the signature rate of SNS-QDS and can be applied to other QDS. More details of the calculation are shown in Appendix~\ref{A3}.
In the simulations, we consider two common cases where each message to be signed is $10^3$ bits (1 Kb) and $10^6$ bits (1 Mb), respectively.  The repetition rate of the laser is 1 GHz, and the distances between Alice-Bob and Alice-Charlie are assumed to be the same. \textcolor{black}{The unit of signature rate is set as time per second (tps).}
Detailed analysis is shown in Appendix~\ref{A4}. Other simulation parameters are listed in Table \ref{tabel2}.

It should be mentioned that all conventional single-bit QDS protocols sign only a one-bit message every round. In the case of signing the multi-bit message, an $m$-bit message must be encoded into a
new sequence with length $h$ by inserting `0' and adding `1' to the original sequence. The signing efficiency, i.e., $\hat{\eta}=m/h$, is obviously less than 1. For simplicity, we use the upper bound $\hat{\eta}=1$, i.e., $h=m$, in our simulation. It is obvious that key consumption of single-bit QDS increases linearly with message size $m$. In other words, the signature rate is proportional to $1/m$. 
In our proposed scheme, the signature is generated by hash functions operating on the message, so that the signature rate is robust against the length of the message. From Eqs. \eqref{eq10} and \eqref{eq11}, $\epsilon$ increases linearly as $m$ increase, but decrease exponentially as $\mathcal{H}_n$ increases. Thus, to guarantee the same epsilon, $\mathcal{H}_n$, which is proportional to group size $n$, increases logarithmically with $m$. Consequently, the signature rate of the proposed scheme is proportional to $-\log_{2}{m}$.

\begin{figure}[t]
	\centering
	\includegraphics[width=8.6cm]{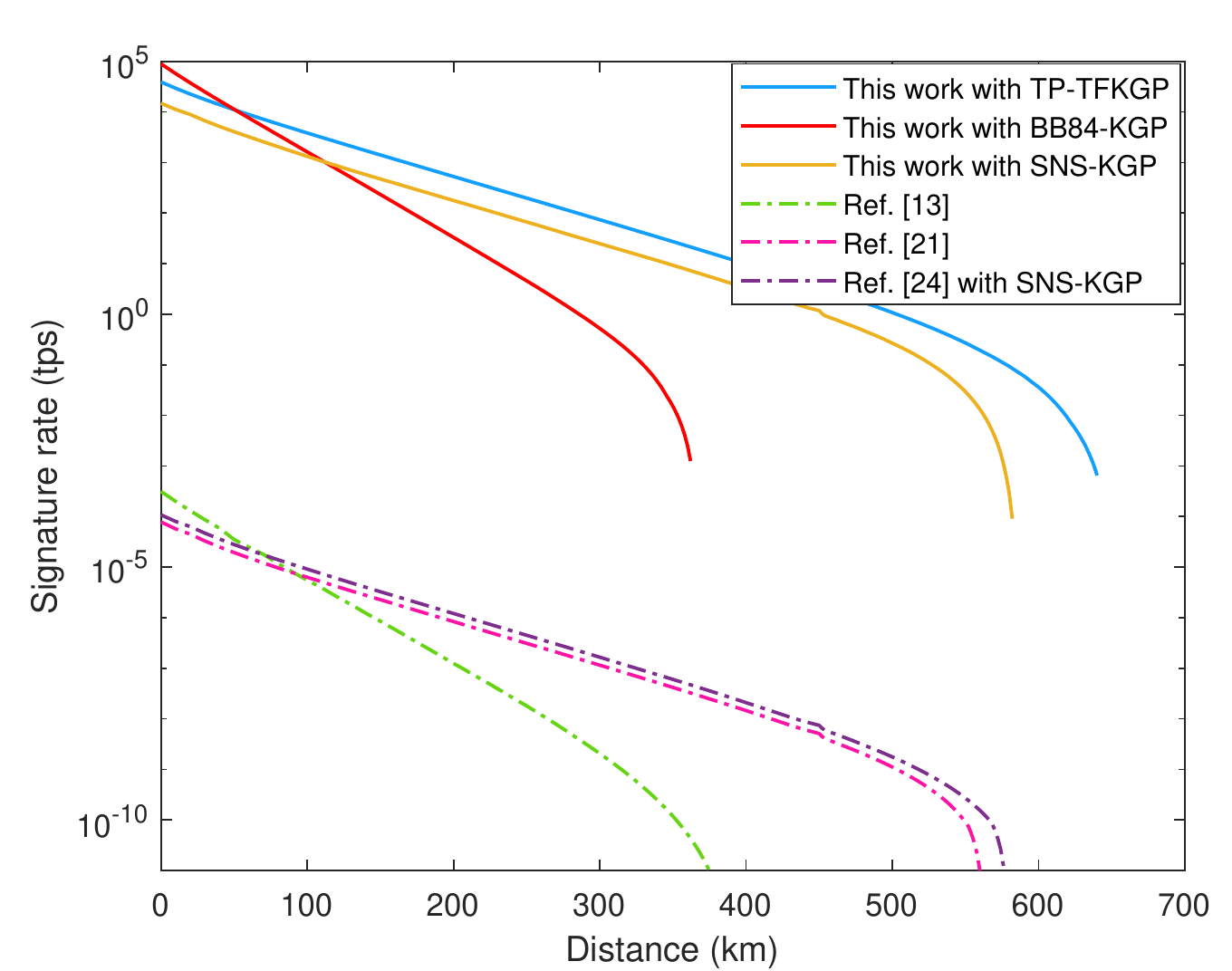}
	\caption{Signature rates of the proposed protocol with TP-TFKGP, BB84-KGP, SNS-KGP, and decoy state BB84-QDS~\cite{Amiri:Secure:2016}, SNS-QDS~\cite{zhang2021twin}, SNS-QDS with random pairing~\cite{qin2022quantum} with the message size of 1 Mb. In the protocol, we use generalized division hashing in messaging stage.
		The repetition rate of the laser is 1 GHz. The distances between Alice--Bob and Alice--Charlie are assumed to be the same. The data size N is $10^{13}$ and the security bound is $10^{-10}$.
	}
	\label{f4}
\end{figure}

\begin{figure}[t]
	\centering
	\includegraphics[width=8.6cm]{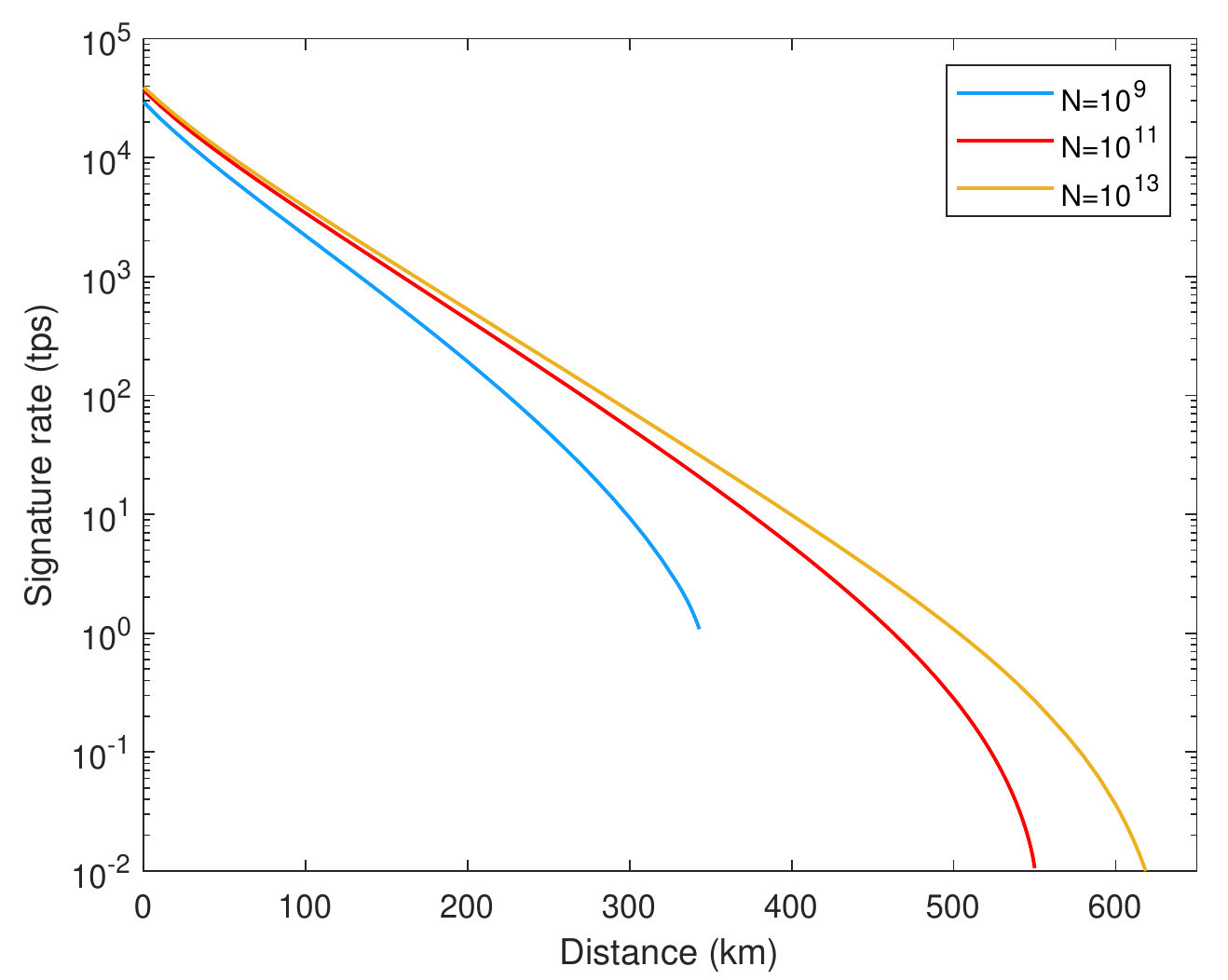}
	\caption{Signature rates of the proposed protocols with TP-TFKGP under different data sizes $N=10^{9},~10^{11}$ and $10^{13}$.  The message size is assumed to be 1 Mb, and the repetition rate of the laser is 1 GHz. The security bound is $10^{-10}$.
	}
	\label{f5}
\end{figure}

The simulation results of all the protocols mentioned are presented in Figs.~\ref{f3} and \ref{f4}. For the message size of 1 Kb,  our protocols show \textcolor{black}{an advantage} on signature rate of over five orders of magnitude compared with conventional QDS schemes, which is a quite larger improvement than SNS-QDS with random pairing. If the message size becomes 1 Mb, the signature rate of conventional BB84-QDS, SNS-QDS, and SNS-QDS with random pairing will decrease by three orders of magnitude, whereas that of our protocols decreases only slightly. Thus the proposed QDS scheme delivers a signature rate \textcolor{black}{with eight orders of magnitude higher than previous schemes.}
As demonstrated, the proposed protocol shows great robustness to message size. Furthermore, based on TP-TFKGP the proposed scheme can reach a transmission distance of 650 km as well as a signature rate of approximately  0.01 times per second (tps).
It is an immense breakthrough  in terms of both distance and signature rate, indicating the considerable potential of the proposed protocol in the practical implementation of QDS.
The performance of the proposed protocol under different data sizes $10^{9}$, $10^{11}$ and $10^{13}$ is depicted in Fig.~\ref{f5}. The curve of $N=10^9$ stops at 1 tps, i.e., one time for all data, because signing less than 1 time (1 message) for all data is nonsense. The result shows that even with a data size as small as $10^9$, the proposed protocol can reach a transmission distance of 350 km\textcolor{black}{, and performance of data size $N=10^{11}$ is close to that of $N=10^{13}$. The influence of finite-size effects caused by small data size on our protocol is in an acceptable level.}

\begin{table*}[tb]
	\centering
	\caption{Time consumption of error correction $T_{\rm EC}$ and privacy amplification $T_{PA}$ under different data sizes $N=10^{13}$ and $N=10^{11}$ when the distance is 400 km. $T_{1}=T_{\rm EC}$ and $T_{2}=T_{\rm EC}+T_{\rm PA}$ represent the postprocessing time of the proposed scheme and OTUH-QDS, respectively. $n_Z$ is the number of raw bits generated in TP-TFKGP; $l$ is the length of keys after privacy amplification. In case $N=10^{13}$, postprocessing time of OTUH-QDS is 5.85 h, and that of the proposed protocol is only 8.07 min.}% The repetition rate of the laser is assumed as 1 GHz}%\label{tab2}
	\begin{tabular}[b]{@{\extracolsep{25pt}}ccccccccc}
		\hline
		\hline
		$N$  & $n_Z$ & errors & l & $T_{\rm EC}$  & $T_{\rm PA}$   & $T_{1}$ & $T_{2}$  \\
		\hline%\xrowht{7pt}
		$10^{13}$ & $1.695\times10^{8}$ & 300 & $4.87\times 10^7$ & 8.07 min  & 5.71 h   & 8.07 min & 5.85 h \\
		%\hline%\xrowht{7pt}
		$10^{11}$ & $1.267\times10^{6}$ & 39830 & $2.51\times 10^5$ & 3.62 s  & 2.98 s   & 3.62 s & 6.6 s \\
		\hline
		\hline
	\end{tabular}
	\label{table3}
\end{table*}

Compared with OTUH-QDS, the proposed protocol does not require perfectly secret keys, and thus involves no privacy amplification step.
Therefore, the proposed protocol only consumes keys with partial information leakage, which is an affordable and practical resource compared with perfect quantum keys generated by quantum secure communication.
Error correction of quantum keys can be easily performed by classical Cascade protocol~\cite{brassard1993secret,yan2008information} where the bit string is first blocked and then manipulated by blocks. Thus the complexity of error correction increases linearly with the data size $N$ and can be performed via stream computing.
Privacy amplification, however, requires a hash matrix multiplication step where the numbers of columns and rows are proportional to $N$. Thus the computational complexity of privacy amplification is $O(N^2)$. The fast Fourier transform algorithm can reduce the complexity to $O(N\log N)$~\cite{hayashi2011exponential}, and one can also block the keys before performing privacy amplification. However, as the minimum blocks should be adequately large to minimize the finite-size effect, the actual computational cost and  delay of privacy amplification are still very large.

The time consumed in conducting consumption of error correction, privacy amplification, and data transmission are listed in Table \ref{table3}, including the total postprocessing time of both protocols, at a distance of 400 km with data sizes $10^{13}$ and $10^{11}$. Details of simulation are introduced in Appendix~\ref{EC}. If $N=10^{11}$, time consumption of postprocessing in OTUH-QDS is 6.6 s, while that of the proposed protocol is 3.62 s. Moreover, when $N=10^{13}$, time for privacy amplification is 5.71 h,  which will 
introduce a quite long delay in experiment. Accordingly, time for error correction is only 8.07 min. The proposed scheme, free of privacy amplification, can significantly save computational resources and minimize postprocessing delays.

\begin{figure}[t]
	\centering
	\includegraphics[width=8.6cm]{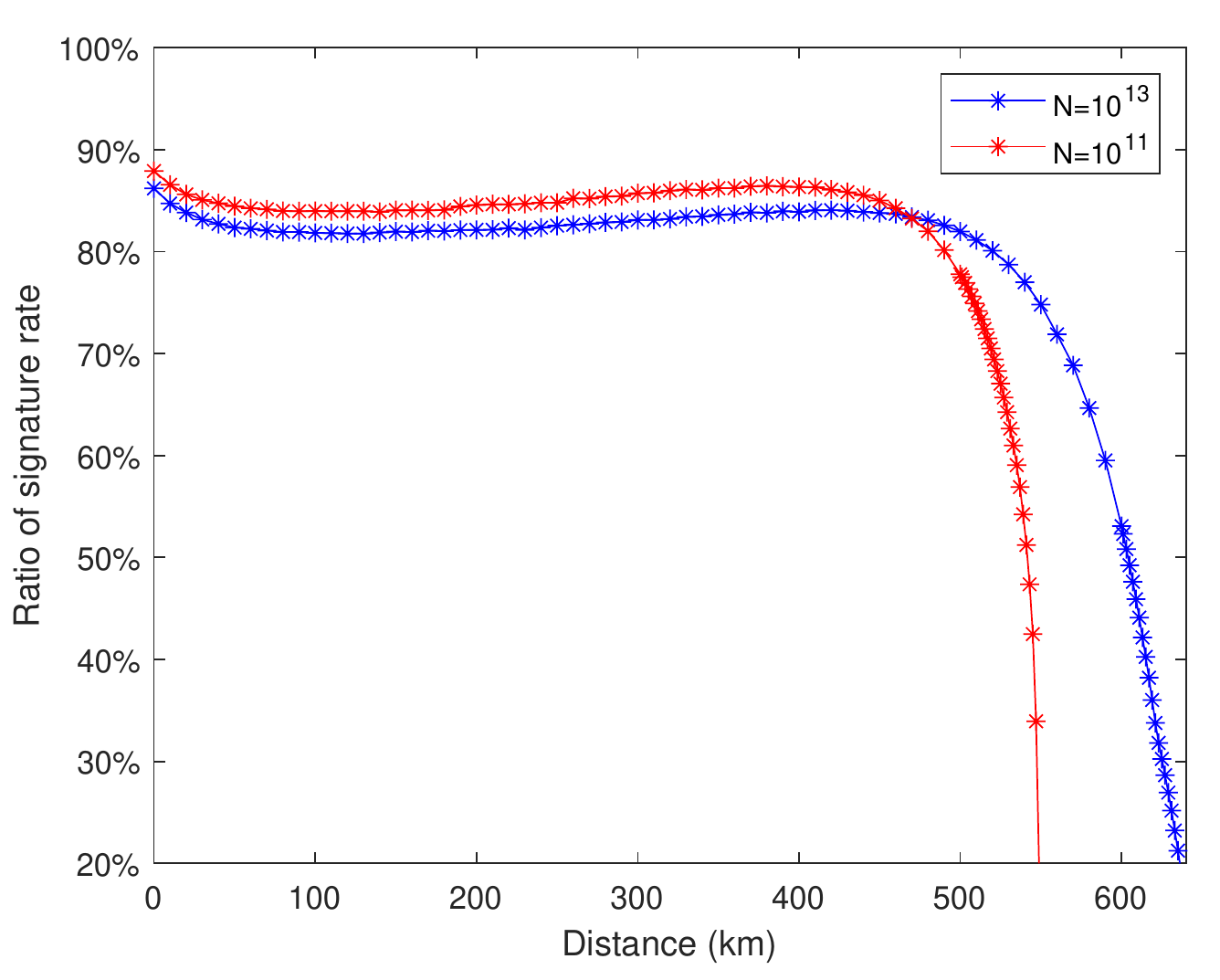}
	\caption{Ratio of the signature rate of the proposed  protocol with TP-TFKGP and that of OTUH-QDS~\cite{yin2023experimental} combined with TP-TFQKD, if not considering the postprocessing time, with data sizes $10^{13}$ and $10^{11}$. The message size is 1 Kb and the repetition rate of the laser is 1 GHz. The ratio is more than 0.8 with a transmission distance lower than 500 km.
	}
	\label{f6}
\end{figure}

We further compare the signature rates of the proposed  protocol and that of OTUH-QDS~\cite{yin2023experimental}.
Theoretically, the two signature rates should be equal under ideal conditions. 
In practical cases, there are two effects that influence the performance of the proposed protocol compared with OTUH-QDS.
The first effect is that in our protocol the parameter $n$ is optimized, which will improve the signature rate compared with OUTH-QDS. This effect will decrease as distance increases. The second effect is that in our protocol we consider the statistical fluctuation of the error rate in the grouping process. This effect will damage the signature rate compared with OUTH-QDS. At both long and short distances, this effect is slight because the size of groups is small and the error rate is small, respectively. 

In Fig.~\ref{f6}, we draw the ratio of the signature rate of the proposed protocol based on TP-TFKGP and that of OTUH-QDS~\cite{yin2023experimental} combined with TP-TFQKD, if not considering the postprocessing time, with data sizes of $10^{13}$ and $10^{11}$,  and message size of 1Kb. 
The result shows that the ratio is more than 80$\%$ for transmission distances  less than  500 km. Overall, the signature rates of the two protocols are comparable. 
In addition, in case of assuming the repetition rate of the laser as 1 GHz, time consumption for postprocessing ($2.057\times10^{4}$s) is even longer than time for data transmission ($10^4$s) for $N=10^{13}$. The signature rate of OTUH-QDS will be constrained by the efficiency of privacy amplification. That is, in practice the signature rate of OTUH-QDS is lower than the simulation result, while the proposed scheme can overcome this shortcoming.
Considering the fact that the proposed protocol can save postprocessing time by even one hundred times, our proposal shows significant \textcolor{black}{improvement} in the practical scenario especially when the digital signature tasks are performed at high frequency and  the data size is large.

\section{Conclusion}\label{conclusion}
		
In summary, in this paper we prove that keys with partial secrecy leakage can protect the authenticity and integrity of messages if combined with AXU hash functions.
Furthermore, we theoretically propose an efficient QDS protocol utilizing imperfect quantum keys without privacy amplification based on the framework of OTUH-QDS, reducing computational resources and  delays of postprocessing without compromising the security. 
The simulation results demonstrate that the proposed protocol outperforms previous single-bit QDS protocols in terms of both signing efficiency and distance.
For instance, for a 1-MB-size message to be signed, the signature rate of the proposed protocol is higher than that of single-bit QDS protocols by over eight orders of magnitude. Specifically, for the protocol based on TP-TFKGP, the transmission distance can reach up to 650 km and still holds a signature rate of 0.01 tps. %In addition, the proposed protocol is robust against security levels and finite-size effects.
Moreover, compared with OUTH-QDS, the proposed protocol notably saves the postprocessing time into an endurable range and therefore, significantly improves the practicality. 
Our scheme is a general framework that can be applied to any existing QKD or QSS protocol, and is highly compatible with future quantum networks and feasible in numerous applications. 
Additionally, this work, only requiring keys with imperfect secrecy, is a new approach of quantum communication that is different from other quantum secret communication protocols.
We suggest that raw quantum keys can be directly used to finish cryptographic tasks including message authentication and digital signatures, indicating the enormous
 potential of this resource  and the possibility of removing the classical postprocessing step in a future quantum world.
We believe that the proposed scheme and the idea of utilizing imperfect quantum keys provide a solution for the real implementation of practical and commercial QDS as well as other quantum cryptography tasks in future quantum networks.

\section*{Acknowledgments}
This study was supported by the National Natural Science Foundation of China (No. 12274223), the Natural Science Foundation of Jiangsu Province (No. BK20211145), the Fundamental Research Funds for the Central Universities (No. 020414380182), the Key Research and Development Program of Nanjing Jiangbei New Area (No. ZDYD20210101),  the Program for Innovative Talents and Entrepreneurs in Jiangsu (No. JSSCRC2021484), and the Program of Song Shan Laboratory (Included in the management of Major  Science and Technology Program of Henan Province) (No. 221100210800-02).

\appendix

\section{single-bit QDS}

\subsection{schematic of single-bit QDS}\label{single-bit QDS}
\begin{figure}[t]
	\centering
	\includegraphics[width=8.6cm]{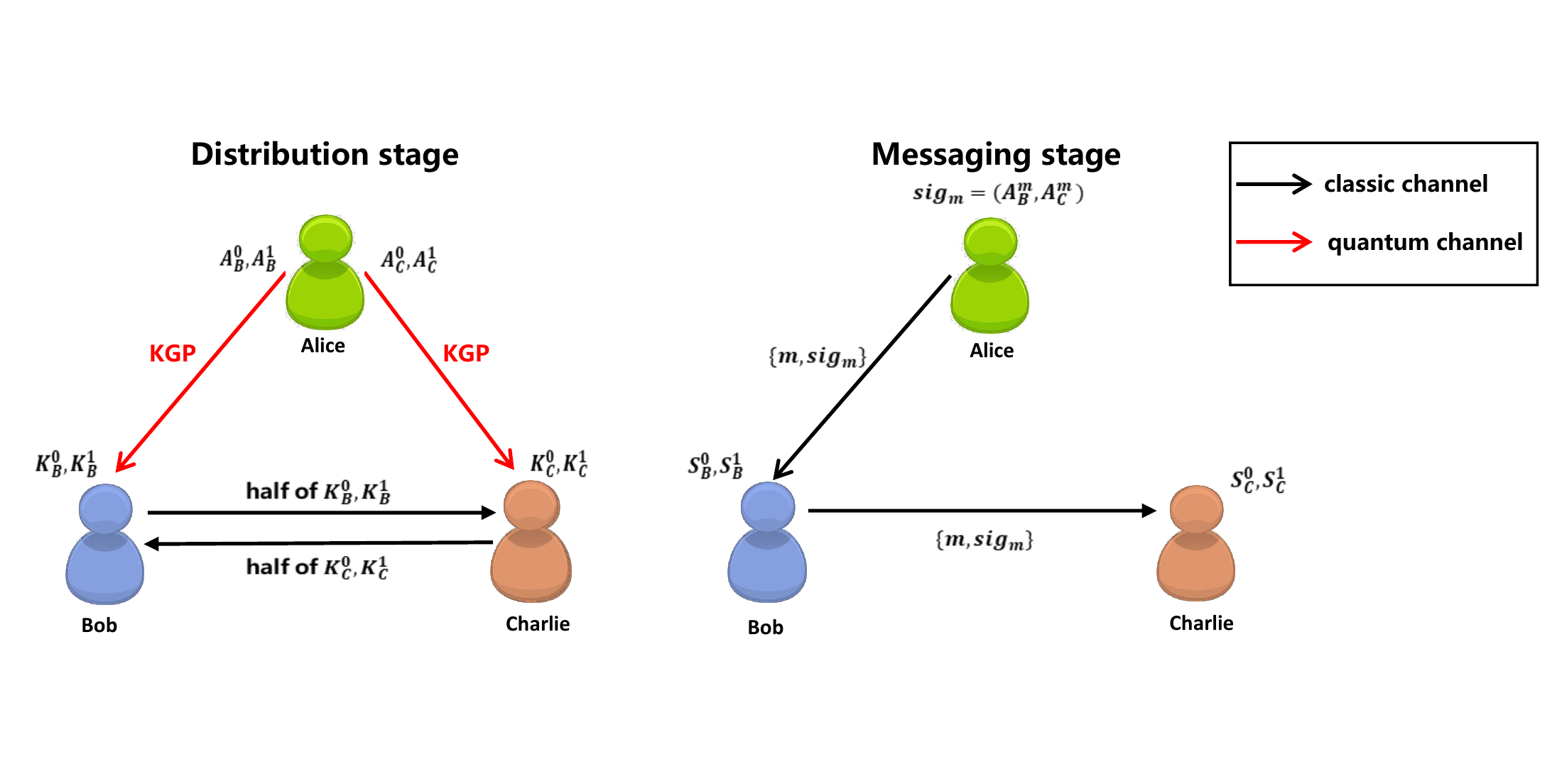}
	\caption{Orthogonal encoding QDS. In the distribution stage, Alice--Bob and Alice--Charlie independently perform KGP to generate correlated bit strings with limited mismatches. Then Bob and Charlie symmetrize their keys by exchanging half of their keys.
		In the messaging stage Alice generates the signature depending on the message bit, and sends the message and signature to Bob, who will transfer it to Charlie.
		Bob and Charlie examine their mismatch and compare it with the threshold to verify the signed message..
	}
	\label{FA1}
\end{figure}

Here, we first introduce orthogonal encoding QDS~\cite{Amiri:Secure:2016} as an example of single-bit QDS schemes. 
Commonly, all single-bit QDS protocols can be segmented into two stages: the distribution stage and messaging stage. 
The schematic of orthogonal encoding QDS is shown in Fig.~\ref{FA1}.

distribution stage:

(i)For each possible future message $m$ = 0 or 1, Alice uses the KGP to generate four different length $L$ keys, $A_B^0,~A_B^1,~A_C^0,~A_C^1$, where the subscript denotes the participant with whom she performed the KGP and the superscript denotes the future message, to be decided later by Alice. Bob holds the length $L$ strings $K_B^0,~K_B^1$ and Charlie holds the length L strings $K_C^0,~K_C^1$.

(ii)For each future message, Bob and Charlie symmetrize their keys by choosing half of the bit values in their $K_B^m,~K_C^m$
and sending them (as well as the corresponding positions) to the other participant using the Bob-Charlie secret classical
channel.
They will only use the bits they did not forward and those received from the other participant. Their final symmetrized keys are denoted as $S_B^m$ and $S_C^m$. Bob (and Charlie) will keep a record of whether an element in $S_B^m$ ($S_C^m$) came directly from Alice or whether it was forwarded to him by Charlie (or Bob).

messaging stage:

(i) To send a signed one-bit message $m$, Alice sends $(m,Sig_m)$ to the desired recipient (say Bob), where $sig_m=(A_B^m,A_C^m)$.

(ii) Bob checks whether $(m,Sig_m)$ matches his $S_B^m$ and records the number of mismatches he finds. He separately checks the part of his key received directly from Alice and the part of the key received from Charlie. If there are fewer than $s_a(L/2)$ mismatches in both halves of the key, where $s_a < 1/2$ is a small threshold determined by the parameters and the desired security level of the protocol, then Bob accepts the message.

(iii) To forward the message to Charlie, Bob forwards the pair $(m,Sig_m)$ that he received from Alice.

(iv) Charlie tests for mismatches in the same way, but in order to protect against repudiation by Alice he uses a different threshold. Charlie accepts the forwarded message if the number of mismatches in both halves of his key is below $s_v(L/2)$ where $s_v$ is another threshold, with $0 < s_a < s_v < 1/2$. 

KGP is actually part of QKD protocol except for the error correction and privacy amplification steps. In distribution stage, $A_X^m$ and $K_X^m$ generated through KGP are correlated with limited mismatch, and $A_X^m$
contains fewer mismatches with $K_X^m$ than does any string produced by an eavesdropper, where $X\in\{B,C\}$ represents Bob and Charlie, $m$ is the message. %Thus forging is forbidden because only Alice knows a valid signature $(A_B^m,A_C^m)$ for a message $m$. 
After Bob and Charlie's symmetrization step, Bob holds $S_B^m$ and Charlie holds $S_C^m$, each containing half of $K_B^m$ and $K_C^m$.
From the perspective of Alice,  $S_B^m$ and $S_C^m$ are symmetric. Alice has no information on whether it is Bob’s $S_B^m$ or Charlie’s $S_C^m$ that contains a particular element of the string ($K_B^m,K_C^m$). This protects against repudiation.
From the perspective of Bob,  $S_B^m$ and $S_C^m$ are asymmetric. Bob has access to all of $K_B^m$ and only half of $K_C^m$, but, even if he is dishonest, he does not know the half of $K_C^m$ that Charlie chose to keep. This protects against forging.

The framework of non-orthogonal encoding QDS is analogous to that of orthogonal encoding. The difference is that it does not require the symmetrization step.  However the signer needs to send the same quantum states to two receivers and only detection events where the two receivers both have clicks are valid.

\subsection{Signing a multi-bit message using single-bit QDS}\label{A4}

The framework above only offer a way for signing a one-bit message. To sign multi-bit messages with these protocols, it is not sufficient to directly iterate the protocol on each bits of the message, which will give a chance for an outside or inside attacker to perform forgery attacks~\cite{wang2015security}. In order to offer information-theoretic security, one must reconstruct the multi-bit message and then sign it bit by bit. This step will make the new message become longer and thus damage the efficiency. To date, the most efficient coding rule is given in Ref.~\cite{zhang2019high}, which can be summarized as follows.

Suppose the signer Alice needs to sign an $n$-bit message $M=m_1||m_2||...||m_n$, $m_i\in \{0,1\}$,~$i=1,2,...,n$. She will encode $M$ into
\begin{equation}
	\begin {aligned}
	    \hat{M}=1_1||1_2||...||1_{x+1}||0||m_1||m_2||...||m_x||0||\\
	    ||m_{x+1}||m_{x+2}||...||m_{2x}||0||\\
	    ...\\
	    ||m_{[\frac{n}{x}]x+1}||m_{[\frac{n}{x}]x+2}||...||m_n||0||1_1||1_2||...||1_{x+1},
    \end{aligned}
\end{equation}
where $x$ refers to the coding interval, $[x]$ is the round down function.
To conclude, the coding rule is that the encoder replenishes a `0' in the head of $M$, and another in the tail. Then,
the encoder inserts a `0' every $x$ bits and  adds `1' with a number of $x + 1$ to both the start and the end.

Denote the length of $\hat{M}$ as $h$. An iteration of conventional QDS protocols with $h$ rounds on $\hat{M}$ is an information-theoretically secure protocol to sign the multi-bit message $M$.
According to the encoding rule, $h=n+[\frac{n}{x}]+2x+4$. For a given $n$ we can optimize $x$ to obtain the minimal $h$ and the maximum efficiency $\eta=\frac{n}{h}$. It is clear that if $n$ is large, the maximum efficiency will be close to 1, but will definitely be less than 1. Thus in our simulation in Sec.~\ref{discussion} we use the upper bound of deficiency, i.e., assume $h=n$.

\section{Mathematical details}

\subsection{Generating an irreducible polynomial}\label{A2}
In this section we introduce ways to generate an irreducible polynomial over Gloise fields GF(2) in random, which is the first step in the messaging stage of our protocol.

Suppose $p(x)$ is a polynomial of order $n$ in ${\rm GF}(2)$. $p(x)$ is irreducible means that no polynomials can divide it except the identity element '1' and $p(x)$ itself.
The necessary and sufficient condition for $p(x)$ being irreducible can be expressed as:
\begin{equation}
	\left\{
	\begin{gathered}
		x^{2^n} \equiv x ~~\rm{mod} ~p(x)
		\\[3pt]
		\rm{gcf}(x^{2^{\frac{n}{d}}}-x,p(x))=1
	\end{gathered}
	\right.
\end{equation}
where $d$ is any prime factor of $n$, $\rm{gcf} ( f(x),g(x) )$ represents the greatest common factor (GCF) of $f(x)$  and $g(x)$.

In order to randomly generate an irreducible polynomial, one way is to generate polynomials at random and test for irreducibility through the condition above. However, this is quite time consuming and requires a lot of random bits.

A better solution is proposed in Ref.~\cite{shoup1996fast}. We can first have an irreducible polynomial of order $n$, defining the extension field GF($2^n$).
Given this, we generate a random element in GF($2^n$) and then compute the minimal polynomial of this element, which will be irreducible.
This procedure only needs $n$ random bits and consumes less time. The concrete procedure is as follows.

Denote the initial irreducible polynomial as $f(x)$ and the polynomial generated by random element as $g(x)$. We will calculate the sequence $a_0=g_0(0)$, $a_1=g_1(0)$, ..., $a_{2n-1}=g_{2n-1}(0)$, where $g_i(x)=g^i(x)$ mod $f(x)$. This sequence of $2n$ elements can fully determine the minimal polynomial of $g(x)$, which can be efficiently computed by Berlekamp-Massey algorithm \cite{massey1969shift}. The result, i.e., the minimal polynomial of $g(x)$, will be the irreducible polynomial we generate.

If choosing generalized division hashing, we need to generate an irreducible polynomial over GF($2^k$). The procedure is the same as that described above. The only difference is that all the calculations need to be done under GF($2^k$).

\subsection{Proof of proposition 1}\label{A22}
An LFSR-based Toeplitz hash function can be  expressed as $h_{p,s}(M)=H_{nm} M$. The construction of $H_{nm}$ is introduced in Def. \ref{def1}. Here we follow the expression in Def. \ref{def1} and define an $n \times n$ matrix $W$ which is only decided by $p$.
\begin{equation}
	W=\begin{pmatrix}
		p_{n-1}& p_{n-2} & ... & p_1 &p_{0} \\
		1 & 0 & ... & 0 & 0 \\
		0 & 1 & ... & 0 & 0 \\
		...&...&...&...&... \\
		0 & 0 & ... & 1 & 0
	\end{pmatrix},
\end{equation}
then we can express $s_i$ through $s$ and $W$
\begin{equation}
	s_i=W^i s.
\end{equation} 
Thereafter we rewrite $h_{p,s}(M)$
\begin{equation}
	\begin{aligned}
	    h_{p,s}(M)=&H_{nm} M \\
	    =&\begin{pmatrix}
	     	s& s_1 & ... &s_{m-1} 
	      \end{pmatrix}
          \begin{pmatrix}
           	M_0\\
           	M_1\\
           	...\\
           	M_{m-1}
          \end{pmatrix} \\
         =&\sum_{i=0}^{m-1} M_i W^i s \\
         =& M(W) s,
    \end{aligned}
\end{equation}
where $M(W)= M_{m-1}W^{m-1}+...+m_1 W +m_0 I $ is an  $n \times n$ matrix.

Define $f(x)$ as the  characteristic polynomial of the matrix $W$, and we can calculate it as follows.
\begin{equation}
	\begin{aligned}
		f(x)=&|xI-W| \\
		=&\begin{vmatrix}
		    x+p_{n-1}& p_{n-2} & ... & p_1 &p_{0} \\
		    1 & x & ... & 0 & 0 \\
		    0 & 1 & ... & 0 & 0 \\
		    ...&...&...&...&... \\
		    0 & 0 & ... & 1 & x 
		  \end{vmatrix} \\
	    =& x^n+p_{n-1}x^{n-1}+...+p_1x+p_0.
	\end{aligned}
\end{equation}
It is obvious that $f(x)=p(x)$, in other words, $p(x)$ is the  characteristic polynomial of the matrix $W$. Then according to Hamilton-Cayley theorem, $p(W)=0$. Thereafter, it is trivial that if $p(x)|M(x)$, $M(W)=0$, and thus $h_{p,s}(M)=M(W) s=0$.

\section{Calculation details}\label{A3}

\subsection{TP-TFQKD and TP-TFKGP in this work}\label{A31}

The calculation of TP-TFKGP in this work is analogous to that in TP-TFQKD~\cite{Xie2023Scalable}.

When Alice and Bob send intensities $k_a$ and $k_b$ with  phase difference $\theta$, the gain corresponding to only one detector ($\mathbf{L}$ or $\mathbf{R}$) clicking is
	\begin{equation}
		\begin{aligned}
			Q_{k_ak_b}^{L\theta}=&y_{k_ak_b}\left(e^{\omega_{k_ak_b}\cos\theta}-y_{k_ak_b}\right),\\
	        Q_{k_ak_b}^{R\theta}=&y_{k_ak_b}\left(e^{-\omega_{k_ak_b}\cos\theta}-y_{k_ak_b}\right).\\
		\end{aligned}
	\end{equation}
where $y_{k_ak_b}=e^{\frac{-(\eta_ak_a+\eta_bk_b)}{2}}(1-p_d)$, $\omega_{k_ak_b}=\sqrt{\eta_ak_a\eta_bk_b}$. The overall gain can be expressed as $Q_{k_ak_b}= 1/2\pi\int_{0}^{2\pi} (Q_{k_ak_b}^{L\theta}+Q_{k_ak_b}^{R\theta})d\theta= 2y_{k_ak_b}[I_0(\omega_{k_ak_b})-y_{k_ak_b}]$, where $I_0(x)$ refers to the zero-order modified Bessel functions of the first kind.

The total number for $\{k_a,k_b\}$ is
\begin{equation}
  x_{k_ak_b}=N p_{k_a}p_{k_b}Q_{k_ak_b}.
\end{equation}

The valid post-matching events on the basis of $Z$ can be divided into two types: correct events $\{\mu_a \mathbf{o}_{a},\mathbf{o}_{b}\mu_b\}$, $\{\mathbf{o}_{a}\mu_a,\mu_b \mathbf{o}_{b}\}$,  and incorrect events $\{\mu_a\mathbf{o}_{a},\mu_b\mathbf{o}_{b}\}$, $\{\mathbf{o}_{a}\mu_a,\mathbf{o}_{b}\mu_b\}$. The corresponding numbers are denoted as
$n_C^z$ and $n_E^z$, respectively, which can be written as
\begin{align}
	n_C^z = x_{\min}\frac{x_{\mathbf{o}_{a} \mu_b}}{x_{0}}\frac{x_{\mu_a\mathbf{o}_{b}}}{x_{1}} =  \frac{x_{\mathbf{o}_{a} \mu_b}x_{\mu_a \mathbf{o}_{b}}}{x_{\max}}\nonumber,
\end{align} 
and 
\begin{align}
	n_E^z = x_{\min}\frac{x_{\mathbf{o}_{a} \mathbf{o}_{b}}}{x_{0}}\frac{x_{\mu_a\mu_b}}{x_{1}}= \frac{x_{\mathbf{o}_{a} \mathbf{o}_{b}} x_{\mu_a \mu_b}}{x_{\max}}\nonumber,
\end{align}
where   $x_{0}=x_{\mathbf{o}_{a}\mu_b}+x_{\mathbf{o}_{a}\mathbf{o}_{b}}$, $x_{1} =x_{\mu_a\mathbf{o}_{b}}+x_{\mu_a\mu_b}$, $x_{\min}=\min\{x_{0},x_{1}\}$, and $x_{\max}=\max\{x_{0},x_{1}\}$.
$s_{11}^z$ corresponds to the number of successful detection events, where Alice and Bob emit a single photon in different time bins in the $Z$ basis. 
The overall number of events in the $Z$ basis is
 	\begin{equation}\label{nz}
 	\begin{aligned}
 		n^z =& n^z_C+n^z_E.	
 	\end{aligned}
 \end{equation}
Considering the misalignment error $e_d^z$, the number of bit errors in the $Z$ basis is $m^z=(1-e_d^z)n^z_E + e_d^z n^z_C$. Thus, the bit error rate in the $Z$ basis is
\begin{equation}\label{Ez}
	\begin{aligned}
		E^z=&\frac{m^z}{n^z}.\\	
	\end{aligned}
\end{equation}

The overall number of ``effective'' events in the $X$ basis is
\begin{equation}
	\begin{aligned}
		n^x&=\frac{1}{\pi}\int_{0}^{\delta} x_{\nu_a\nu_b}^{\theta}d\theta\\
		&=\frac{Np_{\nu_{a}}p_{\nu_{b}}}{\pi}\int_\sigma^{\sigma+\delta} y_{\nu_a\nu_b}(e^{\omega_{\nu_a\nu_b}\cos\theta}\\
		&+e^{-\omega_{\nu_a\nu_b}\cos\theta}-2y_{\nu_a\nu_b})d\theta.\\
	\end{aligned}
\end{equation}
For simplicity, we only consider the case in which all matched events satisfy $\theta^i -\theta^j = 0$. In this case, when $r_a^i \oplus r_a^j \oplus r_b^i \oplus r_b^j=0$ (1), the $\{\nu_a^i\nu_a^j,~\nu_b^i\nu_b^j\}$ event is considered to be an error event when different detectors  (the same detector)  click at time bins $i$ and $j$.

The overall error count in the $X$ basis can be given as
\begin{equation}
	\begin{aligned}
		m^{x}&=\frac{1}{\pi}\int_\sigma^{\sigma+\delta} x_{\nu_a\nu_b}^\theta p_{E}d\theta\\
		&=\frac{2Np_{\nu_{a}}p_{\nu_{b}}}{\pi}\int_\sigma^{\sigma+\delta} y_{\nu_a\nu_b}\times\\
		&\left[\frac{(1-y_{\nu_a\nu_b})^{2}}{e^{\omega_{\nu_a\nu_b}\cos\theta}+e^{-\omega_{\nu_a\nu_b}\cos\theta}-2y_{\nu_a\nu_b}}-1\right]d\theta,\\
	\end{aligned}
\end{equation}		
where $p_{E}= \frac{2q_{\nu_a\nu_b}^{L\theta}q_{\nu_a\nu_b}^{R\theta}}{q_{\nu_a\nu_b}^{\theta}q_{\nu_a\nu_b}^{\theta}}$.

We can then calculate the parameters in Eq. \eqref{1} to estimate the key rate and the information leaked after the distribution stage.
In the following description, let $x^*$ denote the expected value of $x$. We denote the number of $\{k_a,~k_b\}$  as $x_{k_ak_b}$. We denote the number and  error number of events  $\{k_a^{i}k_a^{j},~k_b^{i}k_b^{j}\}$ after post-matching as $n_{k_a^{i}k_a^{j},~k_b^{i}k_b^{j}}$ and $m_{k_a^{i}k_a^{j},~k_b^{i}k_b^{j}}$, respectively.  For simplicity, we abbreviate $k_a^ik_a^j,k_a^ik_a^j$ as $2k_a,2k_b$ when $k_a^i=k_a^j$ and $k_b^i=k_b^j$.

(1)~$\underline{s}_{11}^{z}$.
%The valid post-matching events on the basis of $Z$ can be divided into two types: correct events $\{\mu_a \mathbf{o}_{a},\mathbf{o}_{b}\mu_b\}$, $\{\mathbf{o}_{a}\mu_a,\mu_b \mathbf{o}_{b}\}$,  and incorrect events $\{\mu_a\mathbf{o}_{a},\mu_b\mathbf{o}_{b}\}$, $\{\mathbf{o}_{a}\mu_a,\mathbf{o}_{b}\mu_b\}$. The corresponding numbers are denoted as
%$n_C^z$ and $n_E^z$, respectively, which can be written as

$s_{11}^z$ corresponds to the number of successful detection events, where Alice and Bob emit a single photon in different time bins in the $Z$ basis.
Define $z_{10}$ ($z_{01}$) as the number of events in which Alice (Bob) emits a single photon and Bob (Alice) emits a vacuum state in an $\{\mu_a,\mathbf{o}_b\}$ ($\{\mathbf{o}_a,\mu_b\}$) event. The lower bounds of their expected values are $\underline{z}_{10}^{*}= N p_{\mu_a} p_{\mathbf{o}_{b}} \mu_{a}e^{-\mu_a} \underline{y^*_{10}}$ and $
\underline{z}^*_{01} = N p_{\mathbf{o}_{a}} p_{\mu_b} \mu_{b} e^{-\mu_b} \underline{y^*_{01}} $, respectively, where $\underline{y}_{10}^*$ and $\underline{y}_{01}^*$ are the corresponding yields. These can be estimated using the decoy-state method 
\begin{align}
	\underline{y}_{01}^*\geq& \frac{\mu_b}{N(\mu_b\nu_b-\nu_b^2)} \left(\frac{e^{\nu_b}\underline{x}_{o_a\nu_b}^{*}}{p_{o_a}p_{\nu_b}}\nonumber\right.\\
	&\left.-\frac{\nu_b^2}{\mu_b^2}  \frac{e^{\mu_b}\overline{x}_{\hat{\mathbf{o}}_{a}\mu_b}^{*}}{p_{\hat{\mathbf{o}}_{a}}p_{\mu_b}} - \frac{\mu_b^2-\nu_b^2}{\mu_b^2}\frac{\overline{x}_{o o}^{d*}}{p_{o_ao_b}^d}\right),\\
	\underline{y}_{10}^*\geq&\frac{\mu_a}{N(\mu_a\nu_a-\nu_a^2)} \left( \frac{e^{\nu_a}\underline{x}_{\nu_ao_b}^{*}}{p_{\nu_a}p_{o_{b}}}\nonumber\right.\\
	& \left.-\frac{\nu_a^2}{\mu_a^2} \frac{e^{\mu_a}\overline{x}_{\mu_a\hat{\mathbf{o}}_{b}}^{*}}{p_{\mu_  a}p_{\hat{\mathbf{o}}_{b}}}- \frac{\mu_a^2-\nu_a^2}{\mu_a^2}\frac{\overline{x}_{o o}^{d*}}{p_{o_ao_b}^d}\right),\label{eq_decoy_Y01}
\end{align}
where $x_{oo}^{d}=x_{\hat{\mathbf{o}}_{a}\hat{\mathbf{o}}_{b}}+x_{\hat{\mathbf{o}}_{a}\mathbf{o}_b}+x_{\mathbf{o}_{a}\hat{\mathbf{o}}_{b}}$ represents the number of events where at least one user chooses the declare-vacuum state and $p_{oo}^d=p_{\hat{\mathbf{o}}_{a}}p_{\hat{\mathbf{o}}_{b}}+p_{\hat{\mathbf{o}}_{a}}p_{\mathbf{o}_b}+p_{\mathbf{o}_{a}}p_{\hat{\mathbf{o}}_{b}}$ refers to the corresponding probability.  Thus, the lower bound of $s_{11}^{z*}$ is given by
\begin{equation}
	\begin{aligned}
		\underline{s}_{11}^{z*}= \frac{\underline{z}_{10}^*\underline{z}_{01}^*}{x_{\max}}.\\
	\end{aligned}
\end{equation}

(2)~$\underline{s}_{0\mu_b}^{z}$. $s_{0\mu_b}^z$ represents the number of events in the $Z$ basis, Alice emits a zero-photon state in the two matched time bins, and the total intensity of Bob's pulses is $\mu_b$. We define $z_{00}$ ($z_{0\mu_b}$) as the number of detection events where the state sent by Alice collapses to the vacuum state in the $\{\mu_a,\mathbf{o}_b\}$ ($\{\mu_a,\mu_b\}$) event. The lower bound of the expected values is $\underline{z}_{00}^*={ p_{\mu_a} p_{\mathbf{o}_b}e^{-\mu_a}\underline{x}_{o o}^{d*}}/{p_{o_ao_b}^d}$ and $ ~\underline{z}_{0\mu_b}^*=  {p_{\mu_a} p_{\mu_b}e^{-\mu_a}\underline{x}_{\mathbf{o}_{a}\mu_b}^{*}}/{p_{\mathbf{o}_{a}}p_{\mu_b}}$, respectively. Here, we employ the relationship between the expected value $\underline{x}_{\mathbf{o}_{a}\mu_b}^{*}={ p_{\mathbf{o}_{a}} \underline{x}_{\hat{\mathbf{o}}_{a}\mu_b}^{*}}/{ p_{\hat{\mathbf{o}}_{a}}}$, and $~\underline{x}_{\mathbf{o}_{a}\mathbf{o}_b}^{*}={ p_{\mathbf{o}_{a}} p_{\mathbf{o}_b}\underline{x}_{o o}^{d*}}/{p_{oo}^d}$. The lower bound of $s_{0\mu_b}^{z*}$ can be written as
\begin{equation}
	\begin{aligned}	\underline{s}_{0\mu_b}^{z*}=\frac{\underline{x}_{\mathbf{o}_a\mu_b}^*\underline{z}_{00}^*}{x_{\max}}+\frac{\underline{x}_{\mathbf{o}_a\mathbf{o}_b}^*\underline{z}_{0\mu_b}^*}{x_{\max}}\\
	\end{aligned}
\end{equation}

(3)~$\underline{s}_{11}^{x}$.
We define the phase difference between Alice and Bob as $\theta= \theta_a- \theta_b +\phi_{ab}$.  All valid events in the $X$ basis  can be grouped according to the phase difference $\theta~(\in\{-\delta,\delta\}\cup\{\pi-\delta,\pi+\delta\})$, and the corresponding number in the  $\{k_a,~k_b\}$ event is denoted as $x_{k_ak_b}^\theta $. In the post-matching step, two time bins are matched if they have the same phase difference $\theta$.
Suppose the global phase difference  $\theta$  is a randomly and uniformly distributed value, and considering the angle of misalignment in the  $X$ basis $\sigma$, the expected number of single-photon pairs can be given by
\begin{equation}
	\begin{aligned}
		\underline{s}_{11}^{x*} &=\frac{1}{\pi}\int_\sigma^{\sigma+\delta} x_{\nu_a\nu_b}^\theta\times 2\frac{\nu_be^{-(\nu_a+\nu_b)}\underline{y}_{01}^*}{q_{\nu_a\nu_b}^\theta}\frac{\nu_ae^{-(\nu_a+\nu_b)}\underline{y}_{10}^*}{q_{\nu_a\nu_b}^\theta}d \theta\\
		&=\frac{Np_{\nu_{a}}p_{\nu_{b}}}{\pi}\int_\sigma^{\sigma+\delta}\frac{2\nu_a\nu_be^{-2(\nu_a+\nu_b)}\underline{y}_{01}^*\underline{y}_{10}^*}{q_{\nu_a\nu_b}^\theta},\\
	\end{aligned}
\end{equation}
where $q_{\nu_a \nu_b }^\theta$ is the gain when Alice chooses intensity $\nu_a $, and Bob chooses the intensity $\nu_b$ with phase difference $\theta$ and $x_{\nu_a\nu_b}^{\theta}=N  p_{\nu_a}p_{\nu_b} q_{\nu_a\nu_b}^{\theta}$.

(4)~$\overline{e}_{11}^{x}$.
For single-photon pairs, the expected value of the phase error rate in the $Z$ basis is equal to the expected value of the bit error rate in the $X$ basis. Therefore, we first calculate the number of errors of the single-photon pairs in the $X$ basis ${t_{11}^x}$. The upper bound of ${t_{11}^x}$  can be expressed as
\begin{equation}
	\begin{aligned}
		\overline{t}_{11}^x\leq& m^x - \underline{(m_{\nu_a0,\nu_b0}+m_{0\nu_a,0\nu_b})} +\overline{m}_{00,00}, 	
	\end{aligned}\label{eq_TPTFQKD _t11}
\end{equation}
where  $(m_{\nu_a0,\nu_b0}$ ($m_{0\nu_a,0\nu_b}$) is the error count when the states sent by Alice and Bob in time bin $i$~($j$) both collapse to the vacuum state in events $\{2\nu_a,2\nu_b\}$, and $m_{00,00}$ corresponds to the event where the states sent by Alice and Bob both collapse to vacuum states in events $\{2\nu_a,2\nu_b\}$.  The expected counts $\underline{(n_{\nu_a0,\nu_b0}+n_{0\nu_a,0\nu_b})}^*$ and $\overline{n}_{00, 00}^*$ can be expressed as
\begin{equation}
	\begin{aligned}	
		\underline{(n_{\nu_a0,\nu_b0}+n_{0\nu_a,0\nu_b})}^*=&\frac{2}{\pi}\int_\sigma^{\sigma+\delta} x_{\nu_a\nu_b}^\theta\frac{e^{-(\nu_a+\nu_b)}\underline{q}_{00}^{*}}{q_{ \nu_a\nu_b}^\theta}d\theta\\
		=&\frac{\delta Np_{\nu_{a}}p_{\nu_{b}}e^{-(\nu_a+\nu_b)}\underline{q}_{00}^{*}}{\pi}\\
	\end{aligned}
\end{equation}
and
\begin{equation}
	\begin{aligned}	
		\overline{n}_{00, 00}^*=&\frac{1}{\pi}\int_\sigma^{\sigma+\delta} x_{\nu_a\nu_b}^\theta\left(\frac{e^{-(\nu_a+\nu_b)}\overline{q}_{00}^{*}}{q_{ \nu_a\nu_b}^\theta}\right)^2d\theta\\
		=&\frac{Np_{\nu_{a}}p_{\nu_{b}}}{\pi}\int_\sigma^{\sigma+\delta}\frac{e^{-2(\nu_a+\nu_b)}(\overline{q}_{00}^{*})^2}{q_{\nu_a\nu_b}^\theta}d\theta\\
	\end{aligned}
\end{equation}
respectively. Here $q_{00}^*=x_{o_ao_b}^{d*}/(Np_{oo}^d)$. Using the fact that the error rate of the vacuum state is
always $1/2$, we have $\underline{(m_{\nu_a0,\nu_b0}+m_{0\nu_a,0\nu_b})}^*=\frac{1}{2}\underline{(n_{\nu_a0,\nu_b0}+n_{0\nu_a,0\nu_b})}^*$ and $\overline{m}_{00, 00}^*=\frac{1}{2}\overline{n}_{00, 00}^*$. Hence the upper bound of the bit error rate in the $X$ basis can be given by
\be
\overline{e}_{11}^{x}=\overline{t_{11}^{x}}/\underline{s}_{11}^x.	
\ee

(5)~$\overline{\phi}_{11}^{z}$.
For a failure probability $\varepsilon$, the upper bound of the phase error rate $\phi_{11}^{z}$ can be obtained by using the random sampling without replacement~\cite{yin2020tight}
\begin{equation}
	\begin{aligned}
		\overline{\phi}_{11}^{z}\leq&\overline{e}_{11}^{x}+	\gamma^U \left(\underline{s}_{11}^z,\underline{s}_{11}^x,\overline{e}_{11}^{x},\epsilon\right),\\
	\end{aligned}\label{eq_TPTFQKD _phi11z}
\end{equation}
where
\begin{equation}
	\gamma^{U}(n,k,\lambda,\epsilon)=\frac{\frac{(1-2\lambda)AG}{n+k}+
		\sqrt{\frac{A^2G^2}{(n+k)^2}+4\lambda(1-\lambda)G}}{2+2\frac{A^2G}{(n+k)^2}},
\end{equation}
with $A=\max\{n,k\}$ and $G=\frac{n+k}{nk}\ln{\frac{n+k}{2\pi nk\lambda(1-\lambda)\epsilon^{2}}}$.

(6)~$\underline{s}_{11}^{zn}$,~$\underline{s}_{0\mu_b}^{zn}$ and~$\overline{\phi}_{11}^{zn}$.
Finally we can estimate the parameters in Eq. \eqref{1}, i.e., the lower bound of vacuum events and single-photon pairs in a selected  key group $\underline{s}_{11L}^{z}$ and $\underline{s}_{0\mu_bL}^{z}$, and the upper bound of the phase error rate of the $n$-bit group $\overline{\phi}_{11U}^{z}$. They can be obtained from the parameters above by using the random sampling without replacement.
\begin{equation}
	\begin{aligned}
		\underline{s}_{11}^{zn}\geq&n\left[\underline{s}_{11}^{z}/n^z-	\gamma^U \left(n,n^z-n,\underline{s}_{11}^{z}/n^z,\epsilon\right)\right],\\
		\underline{s}_{0\mu_b}^{zn}\geq&n\left[\underline{s}_{0\mu_b}^{z}/n^z-	\gamma^U \left(n,n^z-n,\underline{s}_{0\mu_b}^{z}/n^z,\epsilon\right)\right],\\
		\overline{\phi}_{11}^{zn}\leq&\overline{\phi}_{11}^{z}+	\gamma^U \left(\underline{s}_{11}^{zn},\underline{s}_{11}^{z}-\underline{s}_{11}^{zn},\overline{\phi}_{11}^{z},\epsilon\right).\\
	\end{aligned}
\end{equation}

(7)$l_{key}$. We can also obtain the length of final keys of TP-TFQKD, which can be used to simulate the performance of OTUH-QDS in Fig.~\ref{f4}.
\begin{equation}\label{l}
	\begin{aligned}
		l_{key} =& \underline{s}_{0\mu_b}^{z}+\underline{s}_{11}^{z}\left[1-H(\overline{\phi}_{11}^{z})\right]-n^zfH(E^z) \\
		&-\log_{2}{\frac{2}{\epsilon_{cor}}}-2\log_{2}{\frac{1}{2\epsilon_{PA}}},
	\end{aligned}
\end{equation}
where $\epsilon_{PA}$ is the failure probability of privacy amplification.

\subsection{BB84-KGP in BB84-QDS and this work}
Both BB84-QDS and this work utilize decoy-state BB84-KGP to generate correlated bit strings.
According to Ref. \cite{lim2014concise} we can estimate the number of vacuum events and single-photon events under $X$ basis,
\be\label{eqn2}
s_{X,0} \geq \tau_{0}\frac{\mu_\rd n_{X,\mu_\rdd}^--\mu_\rdd n_{X,\mu_\rd}^+}{\mu_\rd-\mu_\rdd},
\ee
\begin{multline} \label{eqn3}
s_{X,1} \geq \frac{\tau_{1}\mu_\rs\left[n_{X,\mu_\rd}^--n_{X,\mu_\rdd}^+-\frac{\mu_\rd^2-\mu_\rdd^2}{\mu_\rs^2}(n_{X,\mu_\rs}^+- \frac{s_{X,0}}{\tau_0})\right]}{\mu_\rs(\mu_\rd-\mu_\rdd)-\mu_\rd^2+\mu_\rdd^2}.
\end{multline}
where $\tau_{n}:=\sum_{k\in\cK}e^{-k}k^np_k/n!$ is the probability that Alice sends a $n$-photon state, and
\[
n_{X,k}^\pm:=\frac{e^{k}}{p_k}\left(n_{X,k}\pm\sqrt{ \frac{n_X}{2}\log\frac{21}{\esec}}\right),~\forall~k \in \cK.
\]

We can also calculate the number of vacuum events, $s_{Z,0}$, and the number of single-photon events, $s_{Z,1}$, for $\cZ=\cup_{k\in\cK}\cZ_k$, i.e., by using Eqs.~\eqref{eqn2} and~\eqref{eqn3} with statistics from the basis $Z$.
Then we can obtain the phase error rate of the single-photon events in the $X$ basis by
\be \label{eqn5}
\phi_{X,1}:=\frac{c_{X,1}}{s_{X,1}}  \leq \frac{v_{Z,1}}{s_{Z,1}} + \gamma^{U}\left( s_{Z,1},{s}_{X,1},\frac{v_{Z,1}}{s_{Z,1}},\esec \right), \ee
where
\[
v_{Z,1} \leq \tau_{1}\frac{m_{Z,\mu_\rd}^+-m_{Z,\mu_\rdd}^-}{\mu_\rd-\mu_\rdd},
\]
\[
m_{Z,k}^{\pm}:=\frac{e^{k}}{p_k}\left(m_{Z,k}\pm\sqrt{ \frac{m_Z}{2}\log\frac{21}{\esec}}\right),~\forall~k \in \cK,
\]
\[
\gamma^{U}(n,k,\lambda,\epsilon)=\frac{\frac{(1-2\lambda)AG}{n+k}+
	\sqrt{\frac{A^2G^2}{(n+k)^2}+4\lambda(1-\lambda)G}}{2+2\frac{A^2G}{(n+k)^2}}
\]
%\[
%\gamma\left(a,b,c,d \right):= \sqrt{\frac{(c+d)(1-b)b}{cd\log2}\log_2\left( \frac{c+d}{cd(1-b)b} \frac{21^2}{a^2}\right)}.
%\]

The total number of events under $X$ basis is $n_X=\sum_{k\in\cK} n_{X,k}$
and the number of error events is $m_X=\sum_{k\in\cK} m_{X,k}$.

In BB84-QDS, the unknown information  to the attacker is given by
\be
\mathcal{H}= \underline{s}_{X,0} + \underline{s}_{X,1}(1-h(\overline{\phi}_{X,1})) .
\ee

In our protocol based on BB84-KGP, we need to estimate parameters in a selected $n$-bit group, i.e.,
the lower bound of number of vacuum events and single-photon events under $X$ basis $\underline{s}_{X,0}^n$ and $\underline{s}_{X,1}^n$, and the upper bound of the phase error rate of the single-photon events in the $X$ basis $\overline{\phi}_{X,1}^n$.
\begin{equation}
	\begin{aligned}
		s_{X,0}^n\geq&n\left[\underline{s}_{X,0}/n_Z-	\gamma^U (n,n_Z-n,\underline{s}_{X,0}/n_Z,\epsilon)\right],\\
	\end{aligned}
\end{equation}
\begin{equation}
	\begin{aligned}
		s_{X,1}^n\geq&n\left[\underline{s}_{X,1}/n_Z-	\gamma^U (n,n_Z-n,\underline{s}_{X,1}/n_Z,\epsilon)\right],\\
	\end{aligned}
\end{equation}
\begin{equation}
	\begin{aligned}
		\phi_{X,1}^n\leq&\phi_{X,1}+	\gamma^U \left(\underline{s}_{X,1}^n,\underline{s}_{X,1}-\underline{s}_{X,1}^n,\overline{\phi}_{X,1},\epsilon\right).\\
	\end{aligned}
\end{equation}
Finally we can obtain
\be
\mathcal{H}= \underline{s}_{X,0}^n + \underline{s}_{X,1}^n\left[1-h(\overline{\phi}_{X,1}^n)\right]-\lambda_{EC},
\ee
where $\lambda_{EC}=n h(m_X / n_X)$.

\subsection{SNS-KGP and SNS-QDS with random pairing}
We first follow the calculation in Ref. \cite{jiang2019unconditional}.
Alice and Bob obtain $N_{jk}(jk=\{00,01,02,10,20\})$ instances when Alice sends intensity $j$ and Bob sends state $k$. Here '1' and '2' represent the two intensities used in the KGP. After the sifted step, Alice and Bob obtain $n_{jk}$ one-detector heralded events. We denote the counting rate of source $jk$ as $S_{jk}=n_{jk}/N_{jk}$. With all these definitions, we have
\begin{equation}\label{eq28}
\begin{split}
N_{00}=&\left[(1-p_z)^2p_0^2+2(1-p_z)p_zp_0p_{z0}\right]N,\\
N_{01}=&N_{10}=\left[(1-p_z)^2p_0p_1+(1-p_z)p_zp_{z0}p_1\right]N,\\
N_{02}=&N_{20}=\big[(1-p_z)^2(1-p_0-p_1)p_0\\
&+(1-p_z)p_zp_{z0}(1-p_0-p_1)\big]N.
\end{split}
\end{equation}

In addition, we need to define two new subsets of $X_1$ windows, $C_{\Delta^+}$ and $C_{\Delta^-}$, to estimate the upper bound of $e_1^{ph}$.
The number of instances in $C_{\Delta^\pm}$ is
\begin{equation}\label{eq29}
N_{\Delta^\pm}=\frac{\Delta}{2\pi}(1-p_z)^2p_1^2N.
\end{equation}

We denote the number of effective events of right detectors responding from $C_{\Delta^+}$ as $n_{\Delta^+}^R$, and the number of effective events of left detectors responding from $C_{\Delta^-}$ as $n_{\Delta^-}^L$. And we obtain the counting error rate of $C_{\Delta^{\pm}}$, $T_\Delta=\frac{n_{\Delta^+}^R+n_{\Delta^-}^L}{2N_{\Delta^{\pm}}}$. 

If we denote the expected value of the counting rate of \textbf{untagged} photons as $s_1^{Z*}$, the lower bound of $s_1^{Z*}$ is
\begin{equation}\label{s11l}
\begin{split}
s_1^{Z*}\ge &\underline{s}_1^{Z*}=\frac{1}{2\mu_1\mu_2(\mu_2-\mu_1)}\big[\mu_2^2e^{\mu_1}(\underline{S}_{01}^{*}+\underline{S}_{10}^{*})\\
&-\mu_1^2e^{\mu_2}(\overline{S}_{02}^{*}+\overline{S}_{20}^*)-2(\mu_2^2-\mu_1^2)\overline{S}_{00}^*\big],
\end{split}
\end{equation}
where $S^*_{jk}$ is the expected value of $S_{jk}$, and $\overline{S}^*_{jk}$ and $\underline{S}^*_{jk}$ are the upper bound and lower bound of $S^*_{jk}$ when we estimate the expected value from its observed value.

The expected value of the phase-flip error rate of the \textbf{untagged} photons satisfies
\begin{equation}\label{e11u}
e_1^{ph*}\le \overline{e}_1^{ph*}=\frac{\overline{T}^*_\Delta-\frac{1}{2}e^{-2\mu_1}\underline{S}^*_{00}}{2\mu_1e^{-2\mu_1}\underline{s}_1^{Z*}}.
\end{equation}
Here we use the fact that the error rate of vacuum state is always $\frac{1}{2}$.

If the total transmittance of the experimental setups is $\eta$, then we have
\begin{align*}
n_{00}&=2p_d(1-p_d)N_{00},\\
n_{01}&=n_{10}=2\left[(1-p_d)e^{\eta\mu_1/2}-(1-p_d)^2e^{-\eta\mu_1}\right]N_{01},\\
n_{02}&=n_{20}=2\left[(1-p_d)e^{\eta\mu_2/2}-(1-p_d)^2e^{-\eta\mu_2}\right]N_{02},\\
n_t&=n_{signal}+n_{error},\\
E_z&=\frac{n_{error}}{n_t},\\
n_{\Delta^+}^R&=n_{\Delta^-}^L=\left[T_X(1-2e_d)+e_dS_X\right]N_{\Delta^{\pm}},
\end{align*}

where $N_{00},N_{01},N_{10},N_{02},N_{20},N_{\Delta^{\pm}}$ are defined in Eqs.~\eqref{eq28} and~\eqref{eq29}, and
\begin{align*}
n_{signal}=&4Np_z^2p_{z0}(1-p_{z0})\big[(1-p_d)e^{-\eta\mu_z/2}\\
&-(1-p_d)^2e^{-2\eta\mu_z}\big],\\
n_{error}=&2Np_z^2(1-p_{z0})^2\big[(1-p_d)e^{-\eta\mu_z}I_0(\eta\mu_z)\\
&-(1-p_d)^2e^{-2\eta\mu_z}\big]+2Np_z^2p_{z0}^2p_d(1-p_d),\\
T_X=&\frac{1}{\Delta}\int_{-\frac{\Delta}{2}}^{\frac{\Delta}{2}}(1-p_d)e^{-2\eta\mu_1\cos^2{\frac{\delta}{2}}}d\delta\\
&-(1-p_d)^2e^{-2\eta\mu_1},\\
S_X=&\frac{1}{\Delta}\int_{-\frac{\Delta}{2}}^{\frac{\Delta}{2}}(1-p_d)e^{-2\eta\mu_1\sin^2{\frac{\delta}{2}}}d\delta\\
&-(1-p_d)^2e^{-2\eta\mu_1}+T_X,
\end{align*}
where $I_0(x)$ is the $0$-order hyperbolic Bessel functions of the first kind.

In SNS-QDS, the unknown information to the attacker is given by
\be
\mathcal{H}= \underline{s}_1^{Z*}(1-h(\overline{e}_1^{ph*})) .
\ee

In our protocol based on SNS-KGP, there is
\begin{equation}
	\begin{aligned}
		\underline{s}_{1n}^{Z*}=&n\left[\underline{s}_1^{Z*}-	\gamma^U \left(n,n_Z-n,\underline{s}_1^{Z*}/n_Z,\epsilon\right)\right],\\
		\overline{e}_{1n}^{ph*}=&n\left[\overline{e}_1^{ph*}+	\gamma^U \left(\underline{s}_{1n}^{Z*},\underline{s}_1^{Z*}-\underline{s}_{1n}^{Z*}\,\overline{e}_1^{ph*},\epsilon\right)\right].\\
	\end{aligned}
\end{equation}
and
\be
\mathcal{H}=\underline{s}_{1n}^{Z*}\left[1-h(\overline{e}_{1n}^{ph*})\right] -\lambda_{EC},
\ee
where $\lambda_{EC}=n h(E_z)$.

In SNS-QDS with random pairing, we follow the calculation in Ref.~\cite{qin2022quantum}. After random pairing there are two different phase error rates
\begin{equation}
	\tilde{e}_1'^{ph}=\frac{(e_1^{ph})^2}{(e_1^{ph})^2+(1-e_1^{ph})^2},
\end{equation}
\begin{equation}
	\tilde{e}_2'^{ph}=\frac{1}{2},
\end{equation}
and a new bit flip error rate
\begin{equation}
	E'=2E_z(1-E_z).
\end{equation}
The proportion of untagged bits after random pairing is
\begin{equation}
	\Delta'_{un}=\Delta_{un}^2+2\Delta_{un}(1-\Delta_{un}),
\end{equation}
where $\Delta_{un}=Np_z^2p_{z0}(1-p_{z0})s^Z_1/n_t$ is the proportion of untagged bits before random pairing.
The unknown information to the attacker is given by
\begin{equation}
	\begin{aligned}
		\mathcal{H}=&\Delta'_{un}-\Delta_{un}^2\left[p_1H(\tilde{e}_1'^{ph})+(1-p_1)H(\tilde{e}_2'^{ph})\right]\\
		&-2\Delta_{un}(1-\Delta_{un})H(e_1^{ph}),
	\end{aligned}
\end{equation}
where $p_1=(e_1^{ph})^2+(1-e_1^{ph})^2$.

\section{Error correction and privacy amplification}\label{EC}
In this section we introduce our simulation of error correction and privacy amplification in Table~\ref{table3}. We use the simulated data of TP-TFKGP at the distance of 400 km, which can be calculated  by Eqs.~\ref{nz}, \ref{Ez}, \ref{l} in Appendix \ref{A3}.
We implement our simulation on a desktop computer with an Intel i5-10400 CPU (with RAM of 8 GB).

We use improved Cascade protocol to perform error correction to correct 300 (or 39830) errors among $1.267\times10^6$ (or $1.695\times10^{8}$) bits. The detailed procedure of improved Cascade protocol can be seen in Ref.~\cite{yan2008information}, where users first block their keys, and then perform binary process on each block to correct the errors and start trace-back section to check the error, until there are no errors in each block.
%In our simulation the length of every blocked bit strings is set as $1.267\times10^6$ ($2\times10^6$) bits. 
The results show that time consumption is 3.62 and 930.98 seconds with data sizes $10^{11}$ and $10^{13}$, respectively.

In privacy amplification step, Alice chooses a random universal$_2$ hash function and performs it on the $n_Z$-bit keys after error correction to obtain $l$-bit final keys.  The choice of function is communicated to Bob, who also uses
it to obtain his $l$-bit final keys. In the simulation we utilize a random Toeplitz matrix as the universal$_2$ hash function. When the data size is $10^{13}$, the matrix is too large that it exceeds the storage of our computer. Thus in the algorithm we block the matrix into $10\times10$ (100) submatrices with the same size to accomplish the calculation. For hash manipulations of every submatrix we follow the procedure in Ref.~\cite{tang2019high}, where fast Fourier transform is used to speed up calculation time.  %Because the matrix can be generated beforehand, we only count the time of matrix multiplication.
In the simulation it takes 2.98 and $2.057\times 10^4$ seconds with data sizes $10^{11}$ and $10^{13}$, respectively.

%%%%%%%%%%%%%%%%%%%%%%% References %%%%%%%%%%%%%%%%%%%%%%%%%
% choose a style
%\bibliographystyle{ieeetr}
%\bibliographystyle{unsrt}
%\bibliographystyle{naturemag}
%\bibliographystyle{apsrev}
%\bibliographystyle{apsrev4-2}
%%%%%%%%%%%%%%%%%%%%%%%%%%%%%%%%%%%%%%%
% choose a .bib file
%\bibliography{reftext}
%%%%%%%%%%%%%%%%%%%%%%%%%%%%%%%%%%%%%%


\end{document}